\documentclass{jnmp}
\usepackage{amsmath}
\usepackage{graphicx}
\newcommand{\sech}{\mathrm{sech}\ }

\newcommand{\doublefig}{.4\textwidth}
\newcommand{\smalldoublefig}{.3\textwidth}
\JNMPnumberwithin{equation}{section}
\resetfootnoterule

\newtheorem*{acknowledgments}{Acknowledgments}
\theoremstyle{definition}

\begin{document}
\renewcommand{\evenhead}{J Cuevas, G James, P G Kevrekidis, B A Malomed
and B S\'{a}nchez-Rey}
\renewcommand{\oddhead}{Approximated profiles for discrete solitons in DNLS lattices}

\thispagestyle{empty}

\FirstPageHead{*}{*}{20**}{\pageref{firstpage}--\pageref{lastpage}}{Article}

\copyrightnote{2007}{J Cuevas, G James, P G Kevrekidis, B A Malomed
and B S\'{a}nchez-Rey}

\Name{Approximation of solitons in the discrete NLS equation}

\label{firstpage}

\Author{Jes\'us CUEVAS~$^a$, Guillaume JAMES~$^b$, Panayotis G
KEVREKIDIS~$^c$, Boris A MALOMED~$^d$ and Bernardo
S\'{A}NCHEZ-REY~$^a$.}

\Address{$^a$ Grupo de F{\'i}sica No Lineal, Departamento de
F{\'i}sica Aplicada I, E. U. Polit{\'{e}}cnica, C/ Virgen de
{\'{A}}frica, 7, 41011 Sevilla, Spain. E-mail: jcuevas@us.es, bernardo@us.es\\[10pt]
$^b$ Institut de Math\'ematiques de Toulouse (UMR 5219), INSA de
Toulouse, 135 avenue de Rangueil, 31077 Toulouse Cedex 4, France.
E-mail: Guillaume.James@insa-toulouse.fr\\[10pt]
$^c$ Department of Mathematics and Statistics, University of
Massachusetts, Amherst MA 01003-4515.
E-mail: kevrekid@math.umass.edu\\[10pt]
$^d$ Department of Physical Electronics, Faculty of Engineering, Tel
Aviv University, Tel Aviv 69978, Israel. E-mail:
malomed@eng.tau.ac.il}

\Date{Received Month *, 200*; Accepted in Revised Form Month *, 200*}

\begin{abstract}
\noindent We study four different approximations for finding the
profile of discrete solitons in the one-dimensional Discrete
Nonlinear Schr\"odinger (DNLS) Equation. Three of them are discrete
approximations (namely, a variational approach, an approximation to
homoclinic orbits and a Green-function approach), and the other one
is a quasi-continuum approximation. All the results are compared
with numerical computations.
\end{abstract}

\section{Introduction}

Since the 1960's, a large number of works has focused on the
properties of solitons  in the  Nonlinear Schr\"odinger (NLS)
Equation \cite{Sulem}. As it is well known, the one-dimensional NLS
equation is integrable. Two of the most important discretizations of
this equation admit discrete solitons. One of these discretizations
is known as the Ablowitz-Ladik equation \cite{AL}, which is also
integrable. On the contrary, the other important discretization,
known as the Discrete Nonlinear Schr\"{o}dinger (DNLS) equation, is
not integrable, and discrete soliton solutions must be calculated
numerically. The DNLS equation has many interesting mathematical
properties and physical applications \cite{Panos}. The DNLS equation
models, among others, an array of nonlinear-optical waveguides
\cite{Demetri}, that was originally implemented in an experiment as
a set of parallel ribs made of a semiconductor material (AlGaAs) and
mounted on a common substrate \cite{Silberberg}. It was predicted
\cite{BEC} that the DNLS equation may also serve as a model for
Bose-Einstein condensates (BECs) trapped in a strong optical
lattice, which was confirmed by  experiments \cite{BECexperiment}.
In addition to the direct physical realizations in terms of
nonlinear optics and BECs, the DNLS equation appears as an envelope
equation for a large class of nonlinear lattices (for references,
see \cite{Aubry}, Section 2.4). Accordingly, the
solitons known in the DNLS equation represent intrinsic localized
modes investigated in such chains experimentally \cite{Sievers}
and theoretically \cite{MacKay,PhysToday}. In this context,
previous formal derivations of the
DNLS equation have been mathematically justified for small amplitude
time-periodic solutions in references \cite{James}.

\begin{figure}
\begin{center}
\begin{tabular}{cc}
\includegraphics[width=\doublefig]{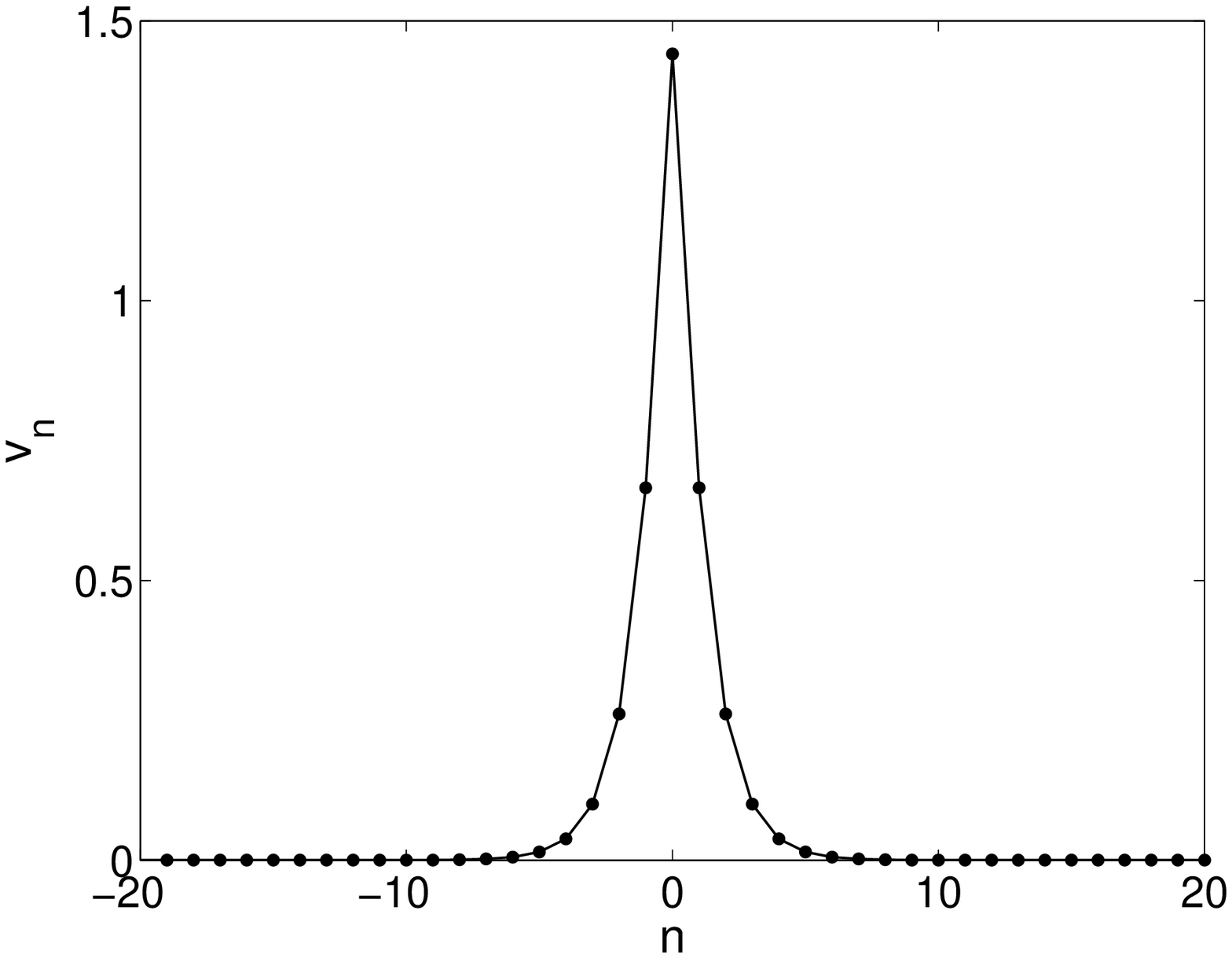} &
\includegraphics[width=\doublefig]{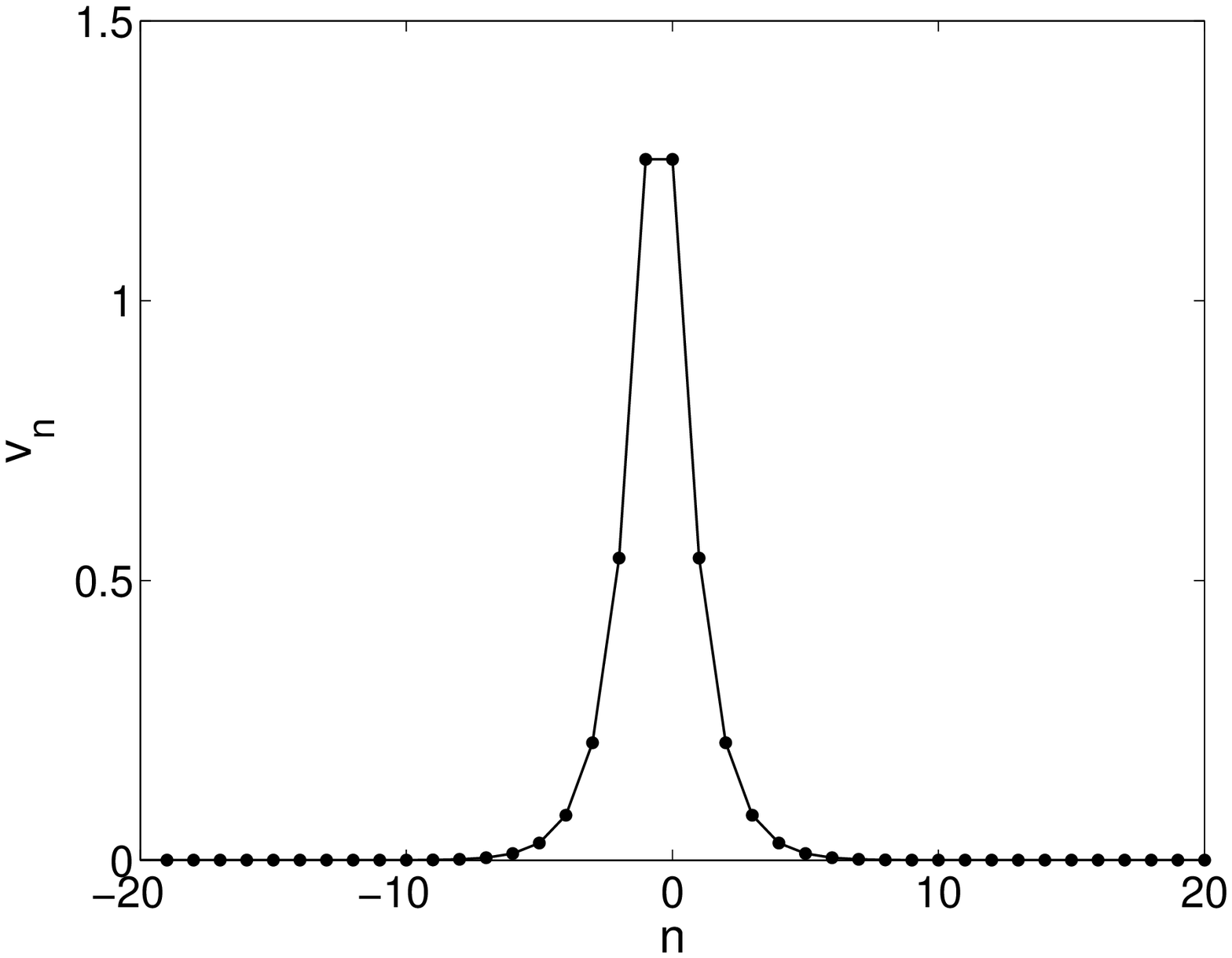}
\end{tabular}%
\end{center}
\caption{Discrete soliton profiles with
$\Lambda=\varepsilon=\gamma=1$. Left panel corresponds to a ST-mode,
and right panel, to a P-mode.} \label{fig:profiles}
\end{figure}

In this paper we will consider fundamental solitons, which are of
two types: Sievers-Takeno (ST) modes, which are site-centered
\cite{ST}, and Page (P) modes, which are bond-centered \cite{Page}
(see also Fig. \ref{fig:profiles}). They can also be seen,
respectively, as discrete solitons with a single excited site, or
two adjacent excited site with the same amplitude. The DNLS equation
is given by
\begin{equation}
i\dot{u}_{n}+\varepsilon \left(
u_{n+1}+u_{n-1}-2u_{n}\right)+\gamma|u_{n}|^{2}u_{n}=0, \label{beq1}
\end{equation}
where $u_{n}(t)$ are the lattice dynamical variables, the overdot
stands for the time derivative, $\epsilon>0$ is the lattice coupling
constant and $\gamma$ a nonlinear parameter.
We look for solutions of frequency $\Lambda$ having the form
$u_{n}(t)=e^{i\Lambda t}v_{n}$.
Their envelope $v_n$ satisfies
\begin{equation}
-\Lambda v_{n}+\varepsilon \left(
v_{n+1}+v_{n-1}-2v_{n}\right)+\gamma|v_{n}|^{2}v_{n}=0. \label{beq2}
\end{equation}
Throughout this paper, we assume $\gamma\varepsilon>0$ and choose
$\gamma=\varepsilon=1$ without loss of generality, as Eq.
(\ref{beq2}) can be rescaled. We also look for unstaggered
solutions, for which, $\Lambda>0$ (staggered solutions with
$\Lambda<0$ can be mapped to the former upon a suitable staggering
transformation $\tilde{v}_n=(-1)^n v_n$). Furthermore, we restrict
to real solutions of (\ref{beq2}), which yield (up to multiplication
by $\exp{i\theta}$) all the homoclinic solutions of (\ref{beq2})
\cite{QX07}. Homoclinic solutions of (\ref{beq2}) can be found
numerically using methods based on the anti-continuous limit
\cite{MacKay} and have been studied in detail (first of all, in
one-dimensional models, but many results have been also obtained for
two- and three-dimensional DNLS lattices) \cite{Panos}.

The aim of this paper is to compare four different analytical
approximations of the profiles of ST- and P-modes together with the
exact numerical solutions. These analytical approximations are of
four types: one of variational kind, another one based on a
polynomial approximation of stable and unstable manifolds for the DNLS map,
another one based on a Green-function method, and, finally, a
quasi-continuum approach.

\section{Discrete approximations}

\subsection{The variational approximation}

\begin{figure}
\begin{center}
\begin{tabular}{cc}
\includegraphics[width=\doublefig]{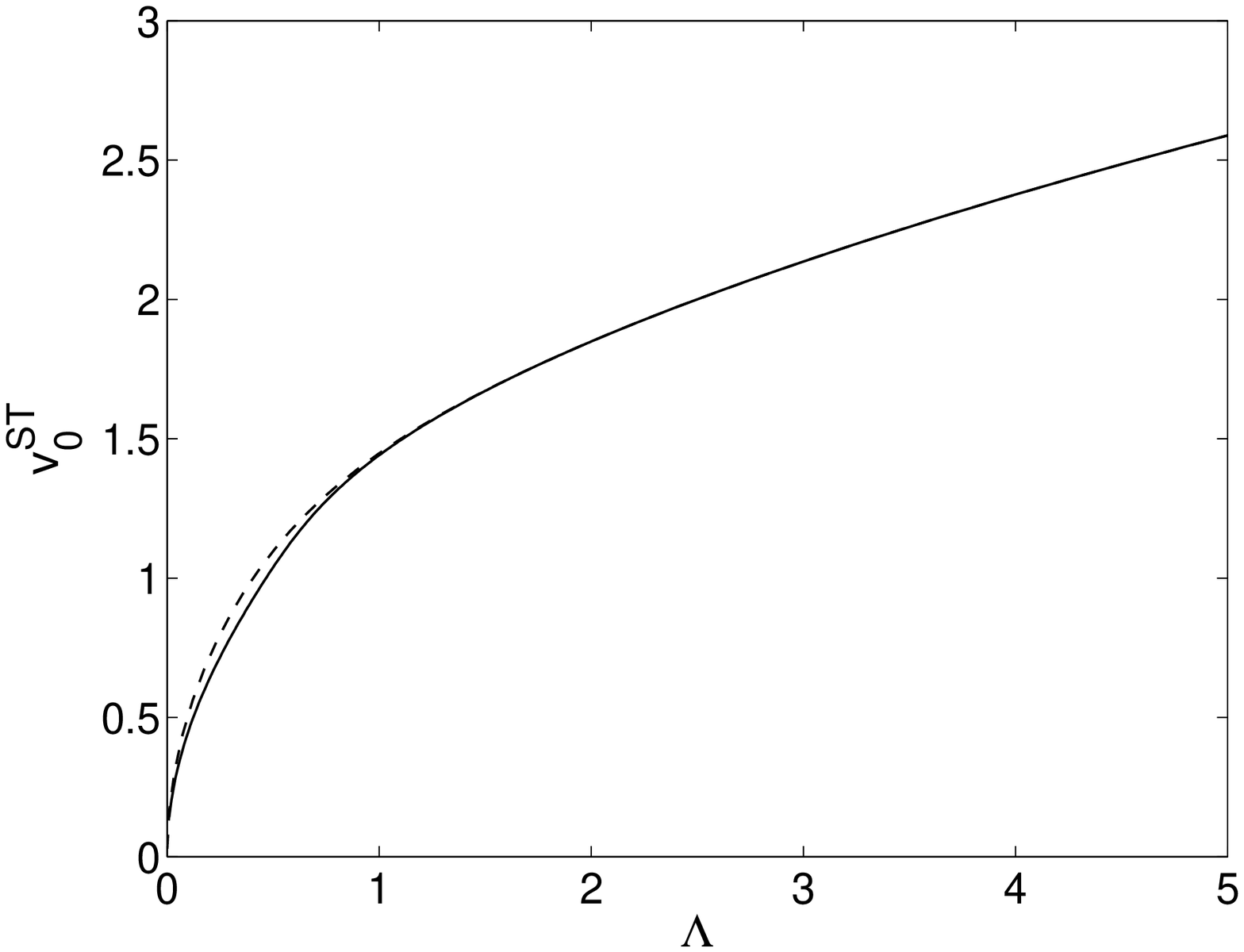} &
\includegraphics[width=\doublefig]{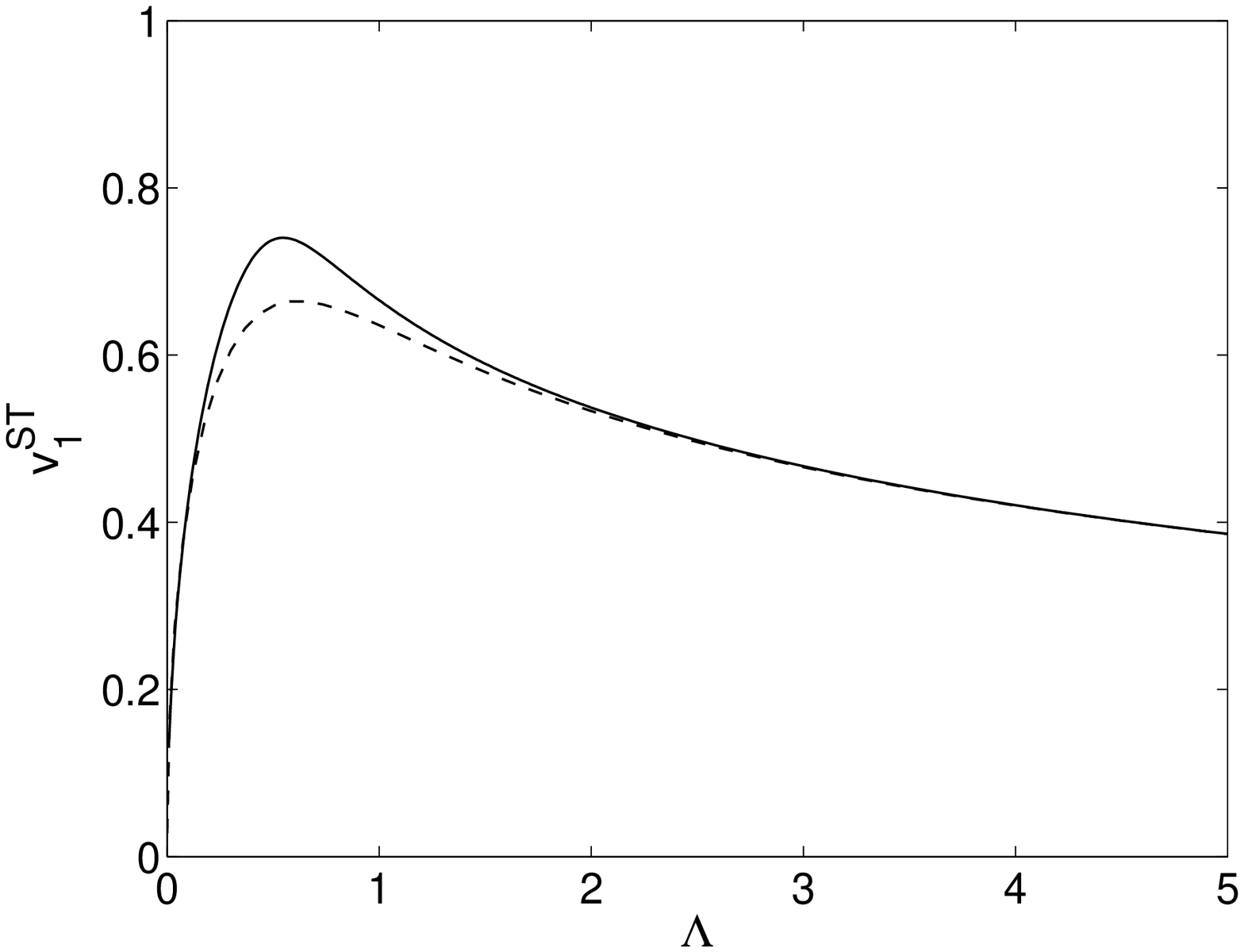}
\end{tabular}%
\end{center}
\caption{Dependence, for ST-modes, of $v_0$ (left panel) and $v_1$
(right panel) with respect to $\Lambda$. Full lines correspond to
the exact numerical solution and dashed lines to the
variational approximation.} \label{fig:profvar1}
\end{figure}

\begin{figure}
\begin{center}
\begin{tabular}{cc}
\includegraphics[width=\doublefig]{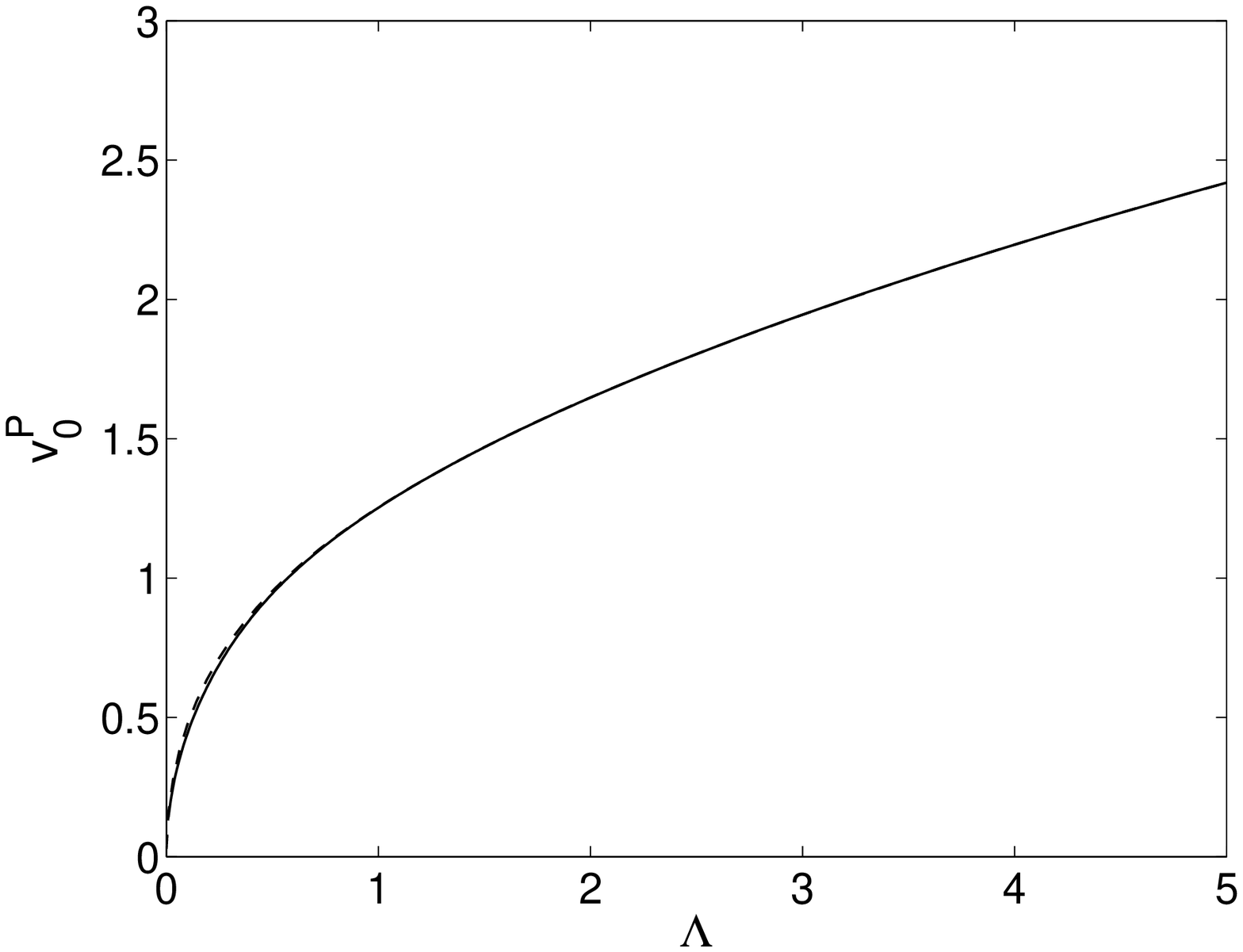} &
\includegraphics[width=\doublefig]{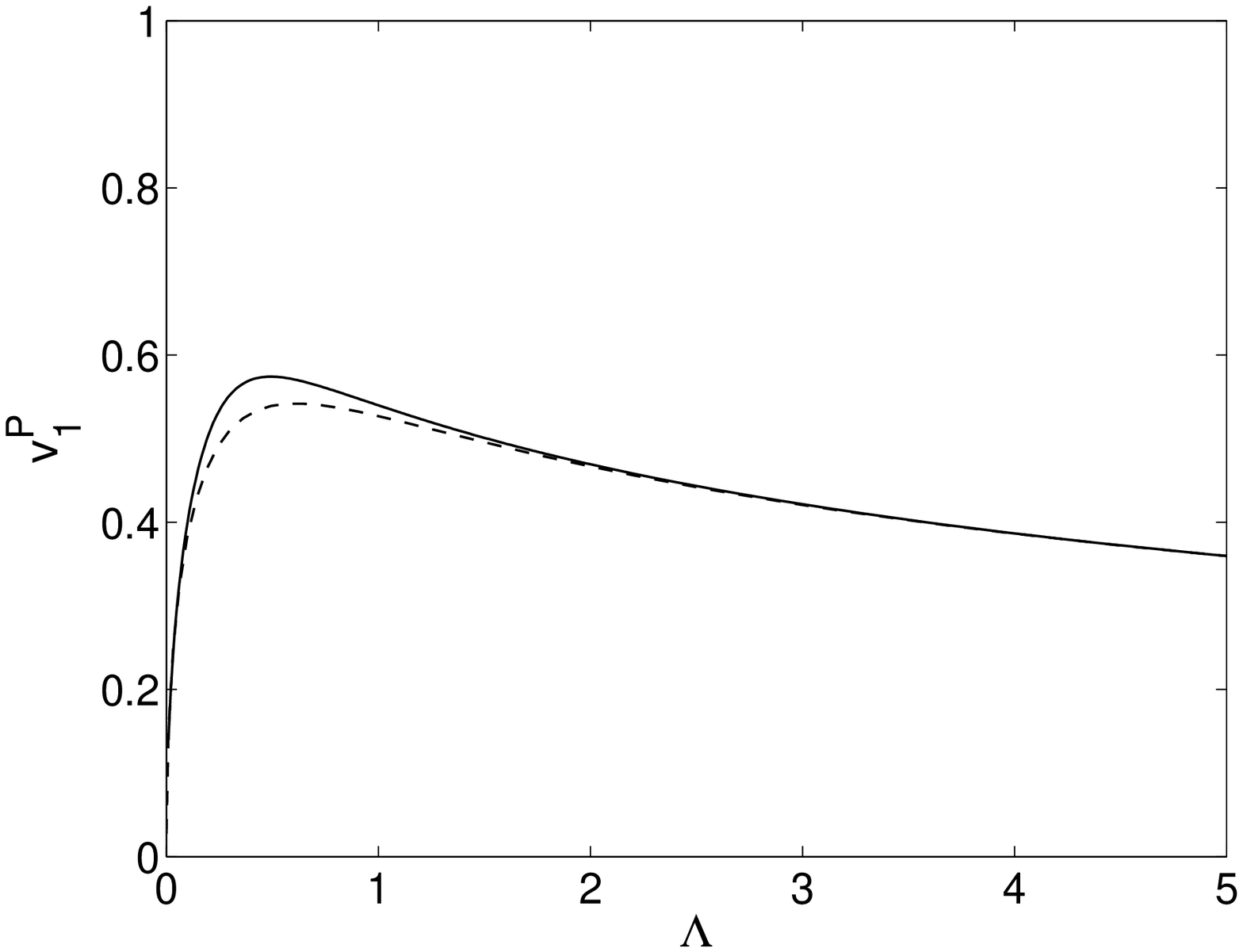}
\end{tabular}%
\end{center}
\caption{Dependence, for P-modes, of $v_0$ (left panel) and $v_1$
(right panel) with respect to $\Lambda$. Full lines correspond to
the exact numerical solution while dashed lines correspond to the
variational approximation.} \label{fig:profvar2}
\end{figure}

Equation (\ref{beq2}) can be derived as the Euler-Lagrange equation for the
Lagrangian
\begin{equation}
L_{\mathrm{eff}}=\sum_{n=-\infty }^{+\infty }\left[
(v_{n+1}+v_{n-1})v_{n}-(\Lambda +2)v_{n}^{2}+\frac{1}{2}%
v_{n}^4\right] .  \label{beq4}
\end{equation}
The VA for fundamental discrete solutions, elaborated in Ref. \cite%
{Weinstein} (see also Ref. \cite{Ricardo}) was based on the simple
exponential ansatz ,
\begin{equation}
v^{ST}_{n}=A_1e^{-a_1|n|},  \qquad v^{P}_{n}=A_2e^{-a_2|n+1/2|},
\label{beq3}
\end{equation}
where $v^{ST}_{n}$ denotes ST-modes, while $v^{P}_{n}$ is for P-modes,
with variational parameters $A_1$, $A_2$, $a_1$ and $a_2$ (which
determine the amplitude and inverse size of the soliton).
Then, substituting the ansatz in the Lagrangian, one can perform the
summation explicitly, which yields the \textit{effective
Lagrangian},
%
%
%
\begin{equation}
L^{ST}_{\mathrm{eff}}=\mathcal{N}_1(2\sech
a_1-\Lambda-2)+\frac{\mathcal{N}_1^2\tanh^2 a_1}{2\tanh 2a_1}, \quad
L^{P}_{\mathrm{eff}}=\mathcal{N}_2\left(\frac{2(1-\cosh a_2)}{\sinh
a_2+\cosh a_2}-\Lambda\right)+\frac{\mathcal{N}_2^2}{4}\tanh
a_2\label{beq5}
\end{equation}
The norm of the ansatz (\ref{beq3}), which appears in Eq.
(\ref{beq5}), is given by $\mathcal{N}\equiv \sum_{n=-\infty
}^{+\infty }v_{n}^{2}$. In particular, for the ST- and P-modes,
\begin{equation}
\mathcal{N}_1=A_1^{2}\coth a_1, \qquad \mathcal{N}_2=A_2^2/\sinh a_2 .
\label{beq7}
\end{equation}
The Lagrangian (\ref{beq5}) gives rise to the variational equations,
$\partial L^{ST}_{\mathrm{eff}}/\partial \mathcal{N}_1=\partial
L^{ST}_{\mathrm{eff}}/\partial a_1=0$, and $\partial
L^{P}_{\mathrm{eff}}/\partial \mathcal{N}_2=\partial
L^{P}_{\mathrm{eff}}/\partial a_2=0$, which constitute the basis of
the VA \cite{Progress}. These predict relations between the norm,
frequency, and width of the discrete solitons within the framework
of the VA, namely
\begin{gather} \label{beq9}
\mathcal{N}_1=\frac{4\cosh a_1\sinh^22a_1}{\sinh4a_1-\sinh2a_1},
\quad \mathcal{N}_2=\frac{8(1-\cosh a_2+\sinh a_2)\cosh^2 a_2}
{\sinh a_2+\cosh a_2}\\ \Lambda=2(\sech
a_1-1)+\mathcal{N}_1\frac{\tanh^2 a_1}{\tanh2 a_1}, \quad
\Lambda=\frac{2(1-\cosh a_2)}{\sinh a_2+\cosh a_2}+
\frac{1}{2}\mathcal{N}_2\tanh a_2 .
\end{gather}
%
%
These analytical predictions, implicitly relating $\mathcal{N}$ and
$\Lambda$ through their parametric dependence on the inverse width
parameter $a$, will be compared with numerical findings below. In
Figs. \ref{fig:profvar1} and \ref{fig:profvar2}, we compare the
approximate and exact values of the highest amplitude site and the
second-highest amplitude sites (i.e. $v_0$ and $v_1$, which can be
easily calculated from (\ref{beq9}) once $\mathcal{N}$ and $a$ are
known) with respect to $\Lambda$ for both ST- and P-modes.
We can observe that the variational approach captures the exact
asymptotic behavior as $\Lambda\rightarrow+\infty$. Indeed as $a_1
\rightarrow+\infty$ in approximation (\ref{beq3})  one obtains
$\Lambda\sim \mathcal{N}_1\sim e^{a_1}$ and $A_1\sim
\sqrt{\mathcal{N}_1}\sim \sqrt{\Lambda}$. Thus  $v^{ST}_0 \sim
\sqrt{\Lambda}$ as $\Lambda\rightarrow+\infty$ which is indeed the
asymptotic behavior of the exact ST-mode. On the contrary, the
variational approximation errs by a small multiplicative factor
($\frac{2}{\sqrt{3}}\sim 1.1$) as $\Lambda\rightarrow 0$ (i.e.,
effectively approaching the continuum limit). This can be seen
taking the limit $a_1 \rightarrow 0$ in approximation (\ref{beq3}).
One has $\mathcal{N}_1\sim 8 a_1$, $\Lambda \sim
-a_1^2+\frac{a_1}{2}\mathcal{N}_1\sim 3a_1^2$ and $A_1\sim
2\sqrt{2}a_1\sim \frac{2}{\sqrt{3}}\sqrt{2\Lambda}$, while the
amplitude of the continuum hyperbolic secant soliton of the
integrable NLS is $A=\sqrt{2 \Lambda}$ [see also below]. Notice that
the P-mode also has the same $\Lambda \rightarrow 0$ limit (and
therefore errs by the same factor).

\subsection{The homoclinic orbit approximation}

\subsubsection{The DNLS map}

\begin{figure}
\begin{center}
\begin{tabular}{cc}
\includegraphics[width=\doublefig]{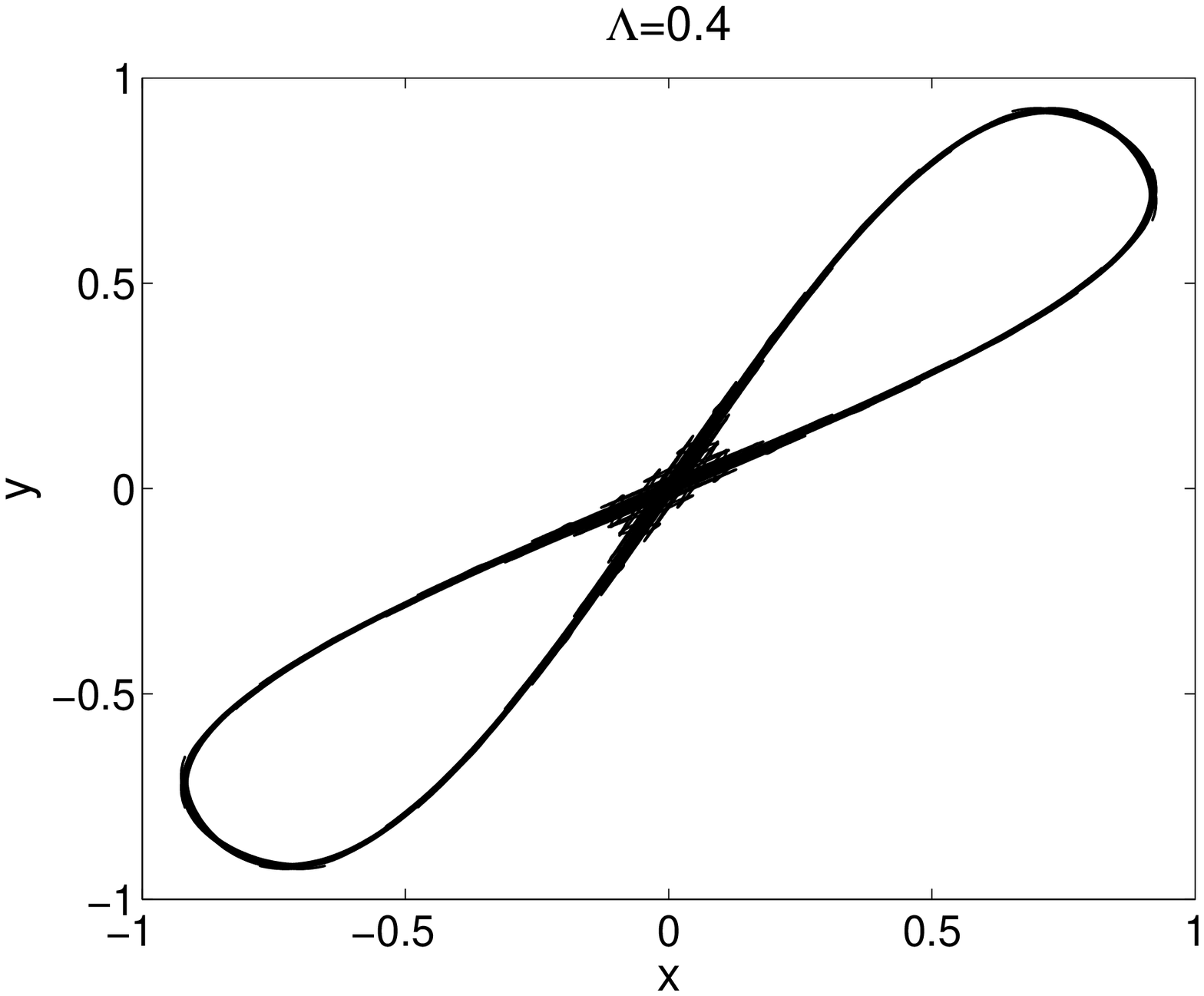} &
\includegraphics[width=\doublefig]{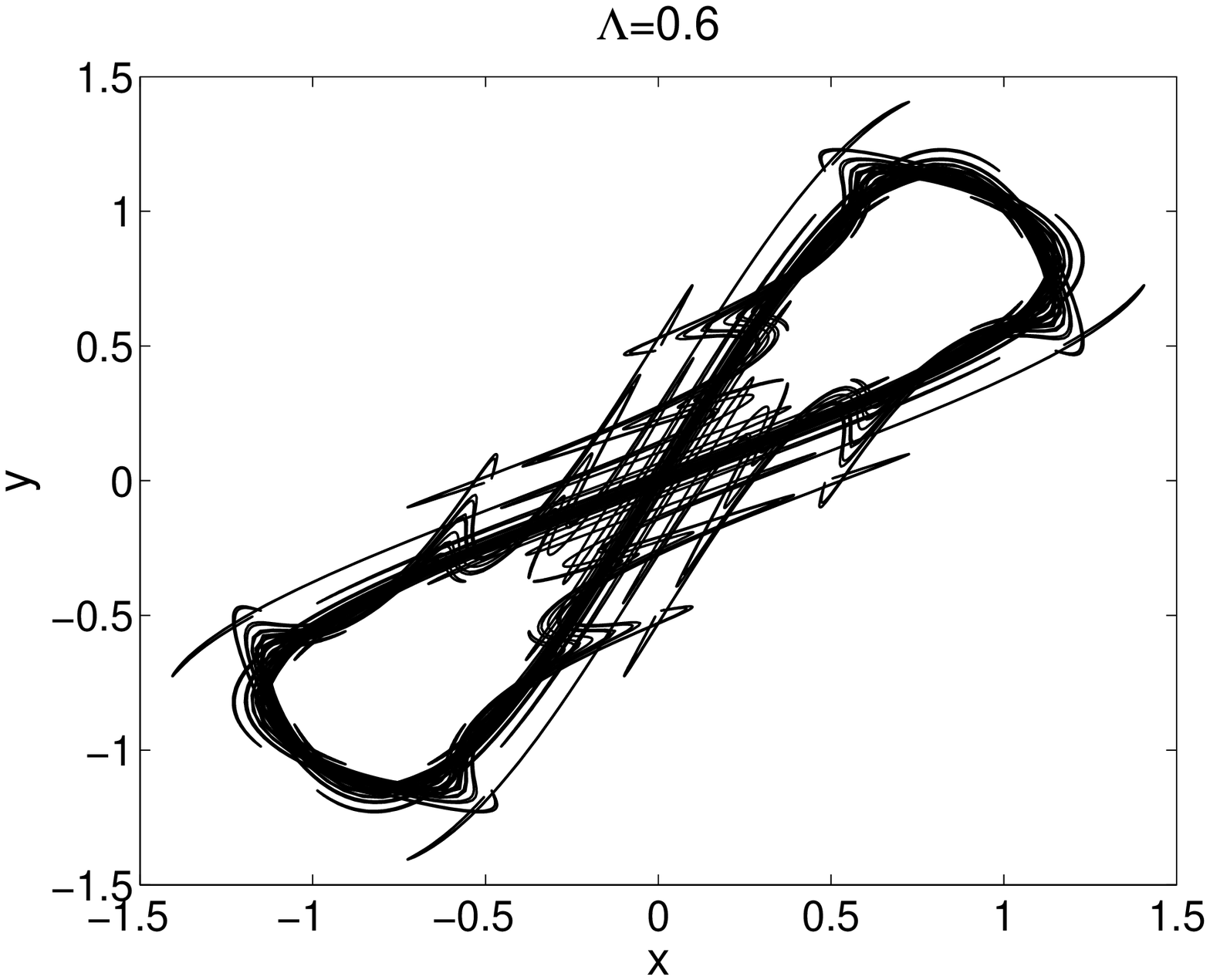} \\
\end{tabular}%
\end{center}
\caption{Homoclinic tangles for $\Lambda=0.4$,
$\Lambda=0.6$.}\label{fig:tangle1}
\end{figure}

\begin{figure}
\begin{center}
\begin{tabular}{cc}
\includegraphics[width=\doublefig]{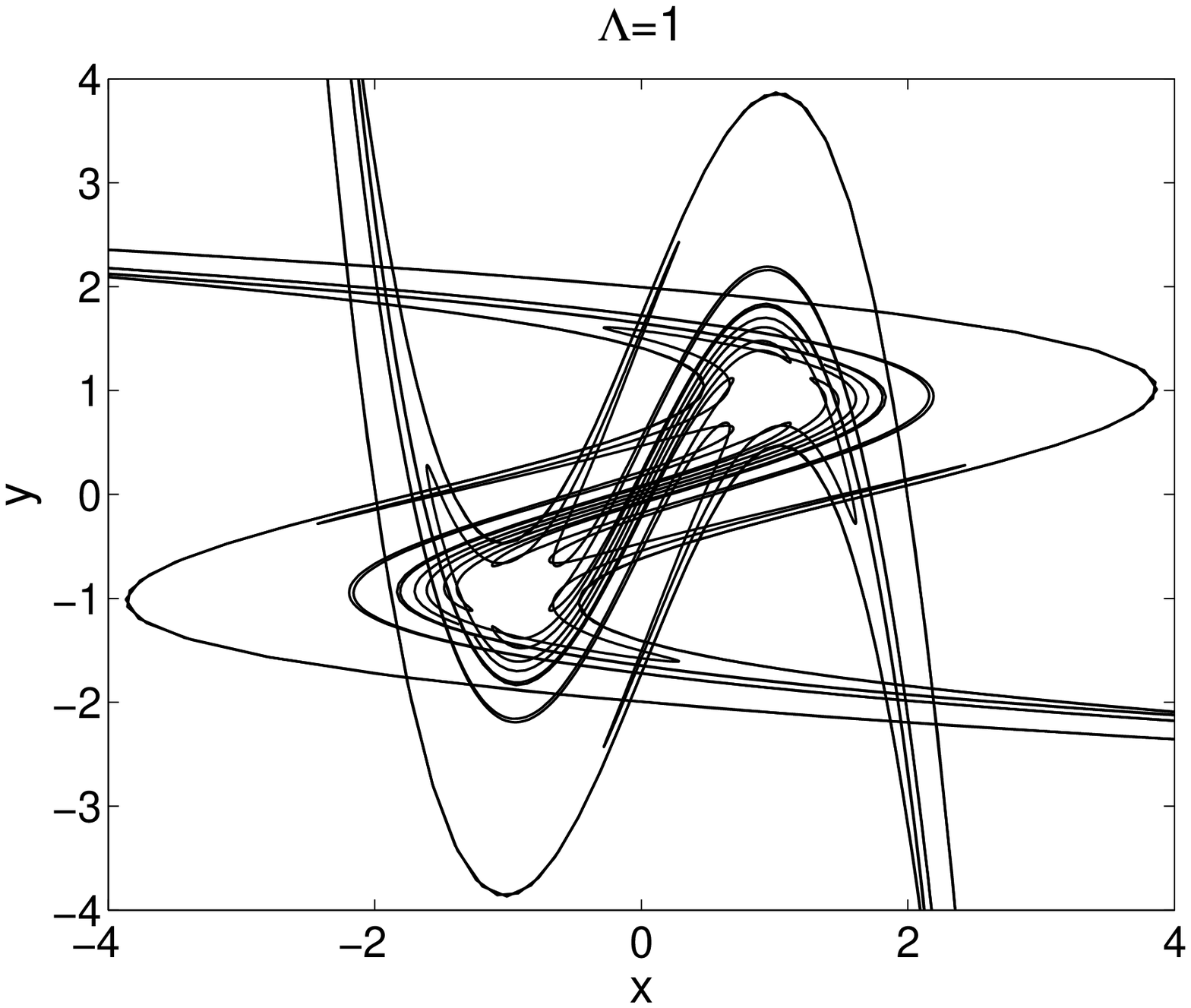} &
\includegraphics[width=\doublefig]{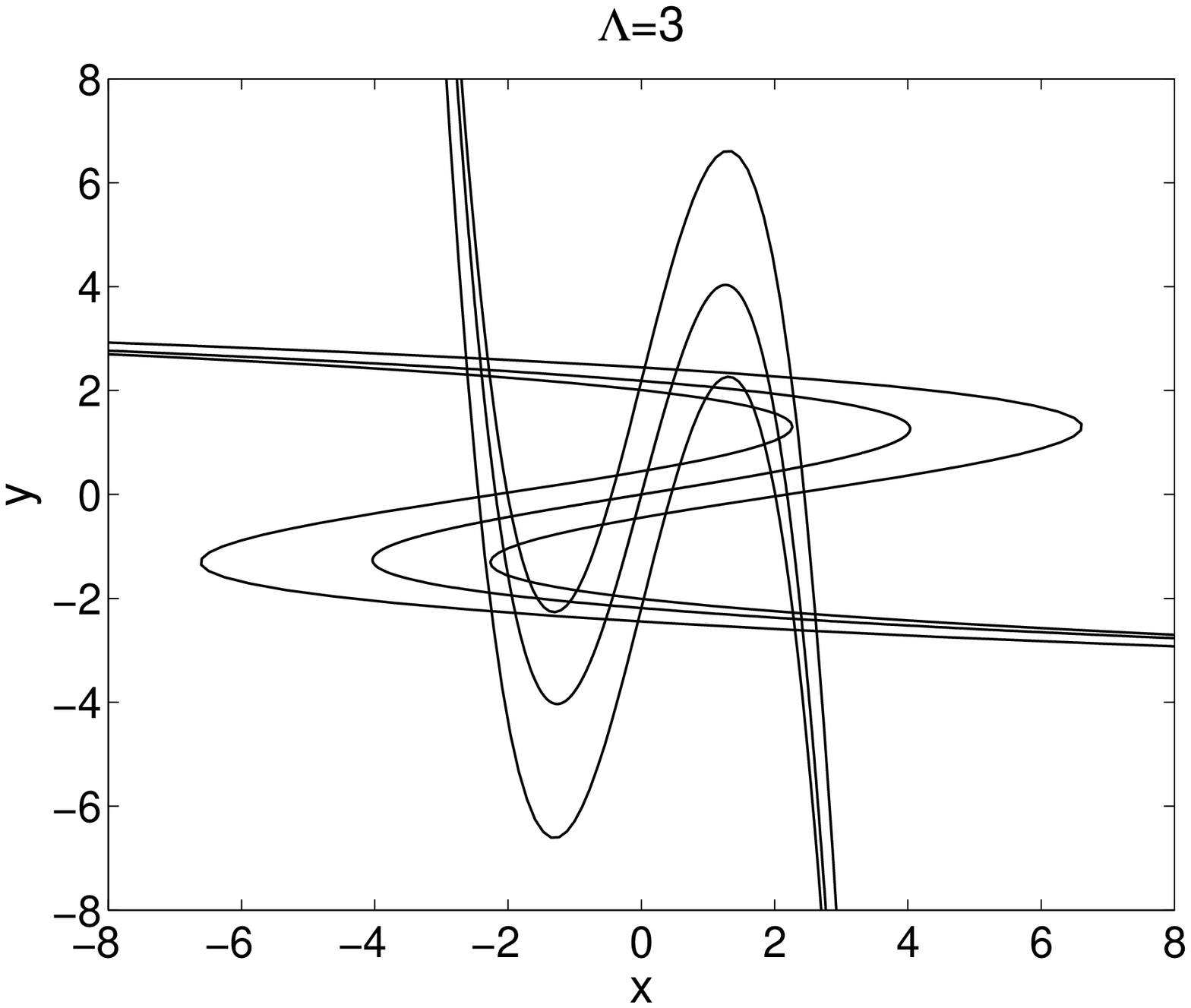}
\end{tabular}%
\end{center}
\caption{Homoclinic tangles for $\Lambda=1$ and $\Lambda=3$.}
\label{fig:tangle2}
\end{figure}

The difference equation (\ref{beq2}) can be recast as a
two-dimensional real map by defining $y_n=v_n$ and $x_n=v_{n-1}$
\cite{Dirk,Bountis,Alfimov,Ricardo,QX07}:
\begin{equation}\label{map}
    \left\{\begin{array}{l}
    x_{n+1}=y_n \\
    y_{n+1}=-y_n^3+(\Lambda+2)y_n-x_n .\\
    \end{array}\right.
\end{equation}
For $\Lambda>0$, the origin $x_n=y_n=0$ is hyperbolic and a saddle
point, which is checked upon linearization of the map around this
point. Consequently, there exists a 1-d stable and a 1-d unstable
manifolds emanating from the origin in two directions given by
$y=\lambda_{\pm}x$, with
\begin{equation}\label{eigen}
    \lambda_{\pm}=\frac{(2+\Lambda)\pm\sqrt{\Lambda(\Lambda+4)}}{2} .
\end{equation}
The eigenvalues $\lambda_{\pm}$ satisfy
$\lambda^2-(\Lambda+2)\lambda+1=0$ and $\lambda_+=\lambda_-^{-1}>1$.
The stable and unstable manifolds are invariant under inversion as
it is the case for eq. (\ref{map}). Moreover, they are exchanged by
the symmetry $(x,y)\rightarrowtail(y,x)$ (this is due to the fact
that the map (\ref{map}) is reversible; see e.g. \cite{QX07} for
more details). Due to the non-integrability of the DNLS equation,
these manifolds intersect in general transversally, yielding the
existence of an infinity of homoclinic orbits (see Figs.
\ref{fig:tangle1} and \ref{fig:tangle2}). Each of their
intersections corresponds to a localized solution, which can be a
fundamental soliton or a multi-peaked one. Fundamental solitons, the
solutions we are interested in, correspond to the primary
intersections points, i.e. those emanating from the first homoclinic
windings. Each intersection point defines an initial condition
$(x_0,y_0)$, that is, $(v_{-1},v_0)$, and the rest of the points
composing the soliton are determined by application of the map.

\subsubsection{The polynomial approximation to the unstable manifold}

\begin{figure}
\begin{center}
\begin{tabular}{cc}
\includegraphics[width=\doublefig]{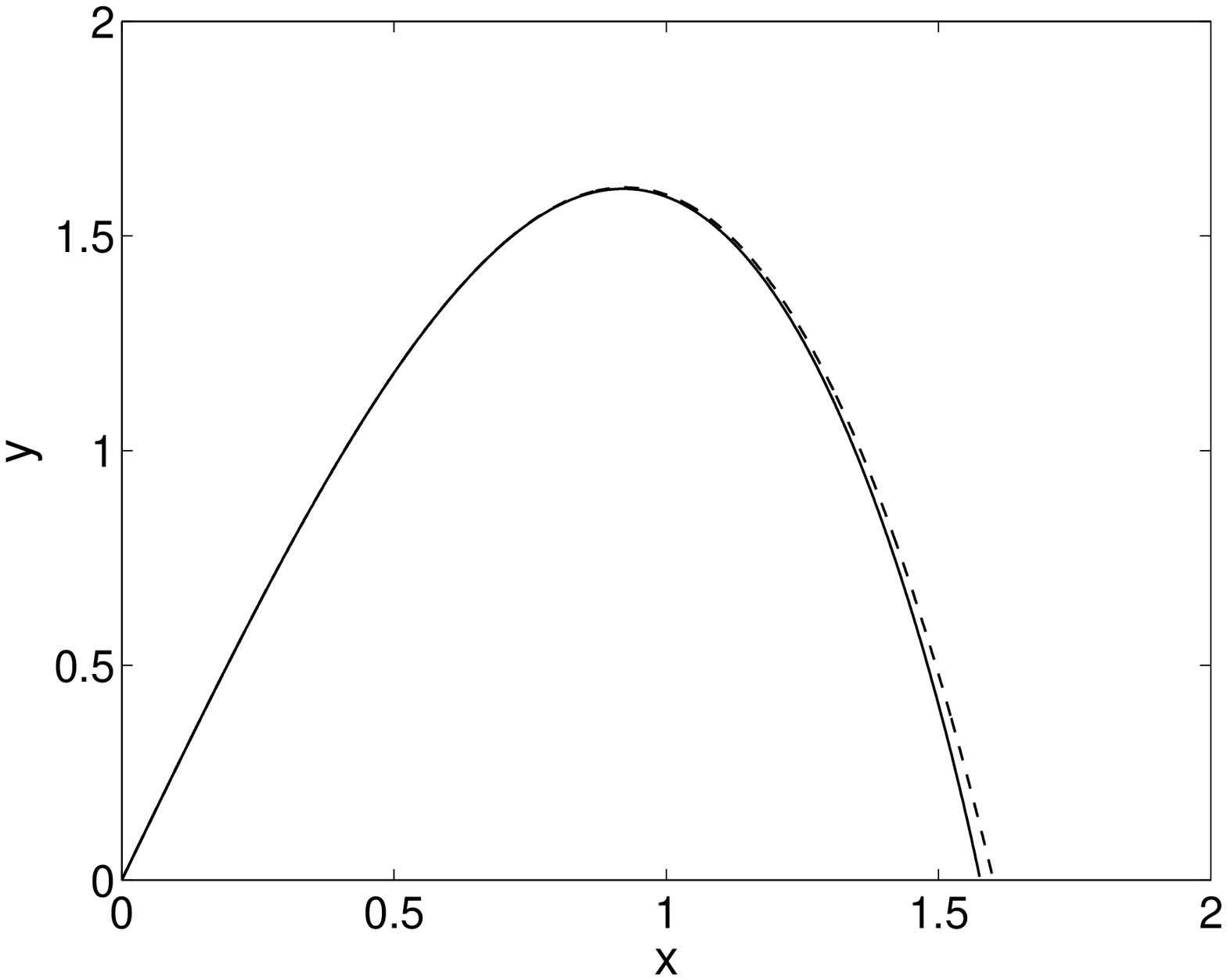} &
\includegraphics[width=\doublefig]{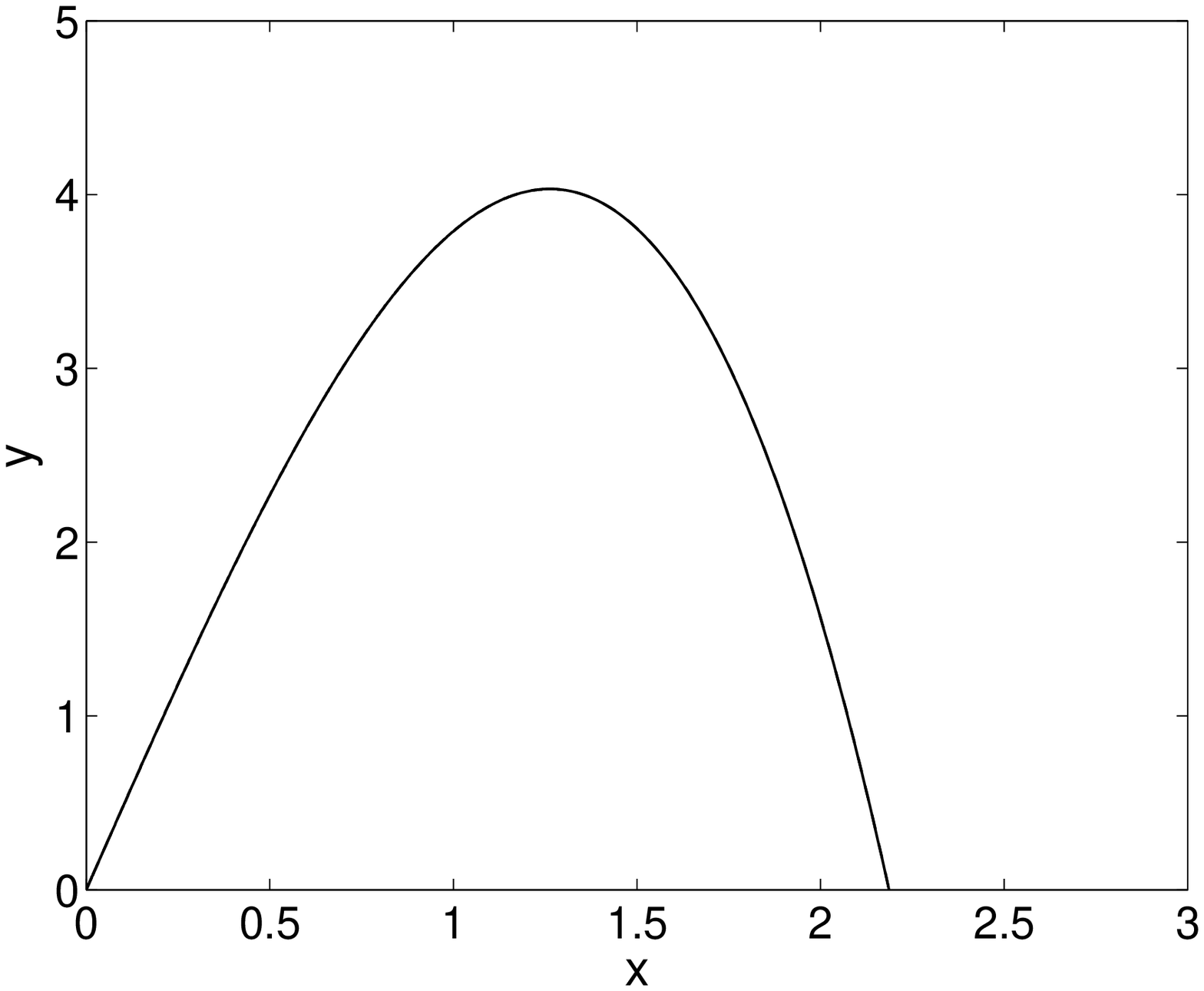}
\end{tabular}%
\end{center}
\caption{Numerical exact unstable manifold
(full line) and its approximation by Eq. (\ref{unstable}) (dashed line)
for $\Lambda=1$ (left panel) and $\Lambda=3$ (right panel). The fit is so accurate in the
latter that both curves are superimposed.} \label{fig:tangleappr}
\end{figure}

The first windings of the stable and unstable manifolds can be
approximated by third order polynomials. Actually, only one of them
is necessary to be determined, as the other one is determined taking
into account the symmetry $x\leftrightarrow y$. We proceed then to
approximate the local unstable manifold $W^u_{\mathrm{loc}}(0)$.
Taking into account its invariance under inversion, it can be
locally written as a graph $y=f(x)=\lambda x-\alpha x^3+O(|x|^5)$
with $\lambda\equiv\lambda_+$ given by (\ref{eigen}). For
$x\approx0$, the image of $(x,f(x))$ under the map (\ref{map}) also
belongs to $W^u_{\mathrm{loc}}(0)$, thus $
-f(x)^3+(\Lambda+2)f(x)-x=f(f(x))\ \forall x\approx0$. This yields
$[\lambda^3+\alpha(\Lambda+2-\lambda-\lambda^3)]x^3+O(|x|^5)=0,\
\forall x\approx0$. Hence
$\alpha=-\lambda^3/(\Lambda+2-\lambda-\lambda^3)=\lambda^4/(\lambda^4-1)$.
The local unstable manifold is approximated at order 3 by
\begin{equation}\label{unstable}
    W^u: y=\lambda x-\frac{\lambda^4}{\lambda^4-1} x^3,
\end{equation}
and, by symmetry, the stable manifold is approximated by:
\begin{equation}\label{stable}
    W^s: x=\lambda y-\frac{\lambda^4}{\lambda^4-1} y^3.
\end{equation}
In Fig. \ref{fig:tangleappr}, the numerical and approximated
unstable manifolds for $\Lambda=1$ and $\Lambda=3$ are compared. It
can be observed that the fit is better when $\Lambda$ increases. The
approximation breaks down for small $\Lambda$ because the origin is
not a hyperbolic fixed point for $\Lambda=0$.

\begin{figure}
\begin{center}
\begin{tabular}{cc}
\includegraphics[width=\doublefig]{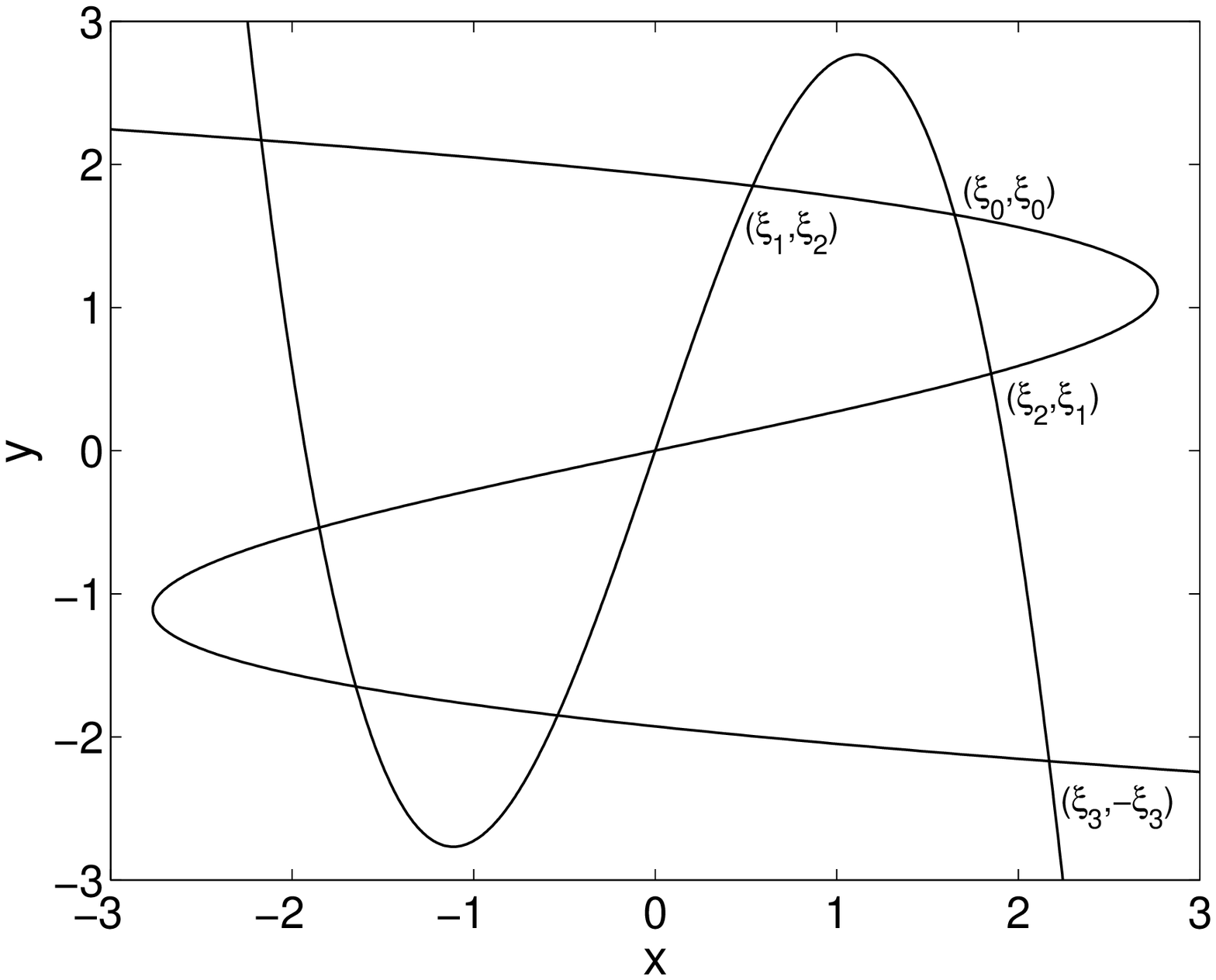} &
\includegraphics[width=\doublefig]{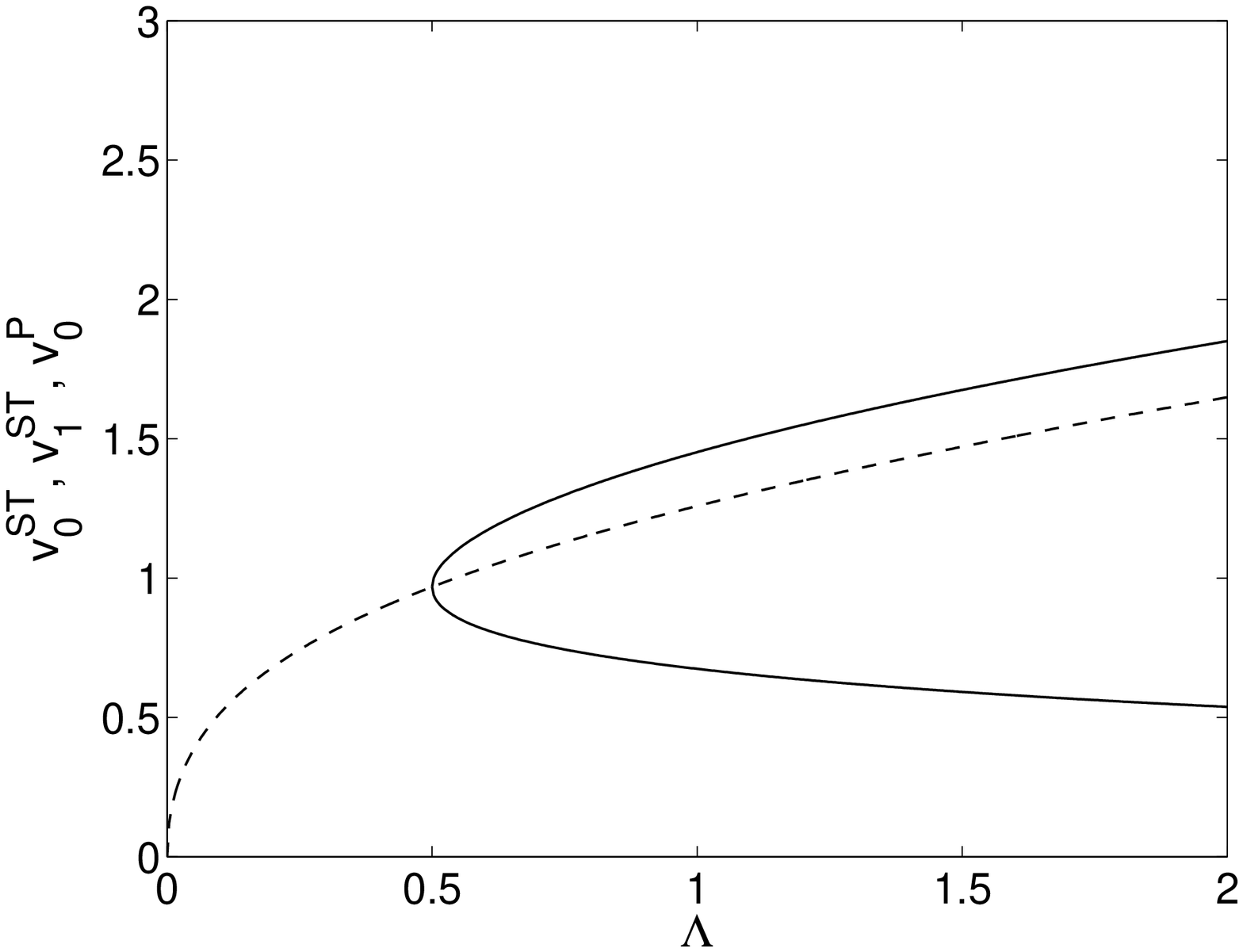}
\end{tabular}%
\end{center}
\caption{(Left panel) Approximated stable and unstable manifolds for
$\Lambda=2$ showing the main intersections. (Right panel) Pitchfork
bifurcation arising in the homoclinic approximation when $\Lambda$
is varied. ST-modes (full lines) bifurcate with the P-mode (dashed
line) at $\Lambda=0.5$.} \label{fig:intersections}
\end{figure}

\subsubsection{Approximate solutions via approximate invariant
manifolds}

Once an analytical form of the unstable and stable manifold is
found, discrete solitons profiles (or, concretely, $v_0$ and
$v_{-1}$) can be determined as the intersection of both manifolds.
The polynomial form of (\ref{unstable}) is not sufficient in
practice to obtain good approximations of the whole soliton profile,
due to sensitivity under initial conditions. However, it provides a
good approximation near the soliton center. Some intersections of
$W^s$ and $W^u$ can be approximated by:
\begin{equation}\label{bigeq}
    x=\lambda\left(\lambda x-\frac{\lambda^4}{\lambda^4-1} x^3\right)-
    \frac{\lambda^4}{\lambda^4-1}\left(\lambda x-\frac{\lambda^4}{\lambda^4-1} x^3\right)^3.
\end{equation}
\begin{figure}
\begin{center}
\begin{tabular}{cc}
\includegraphics[width=\doublefig]{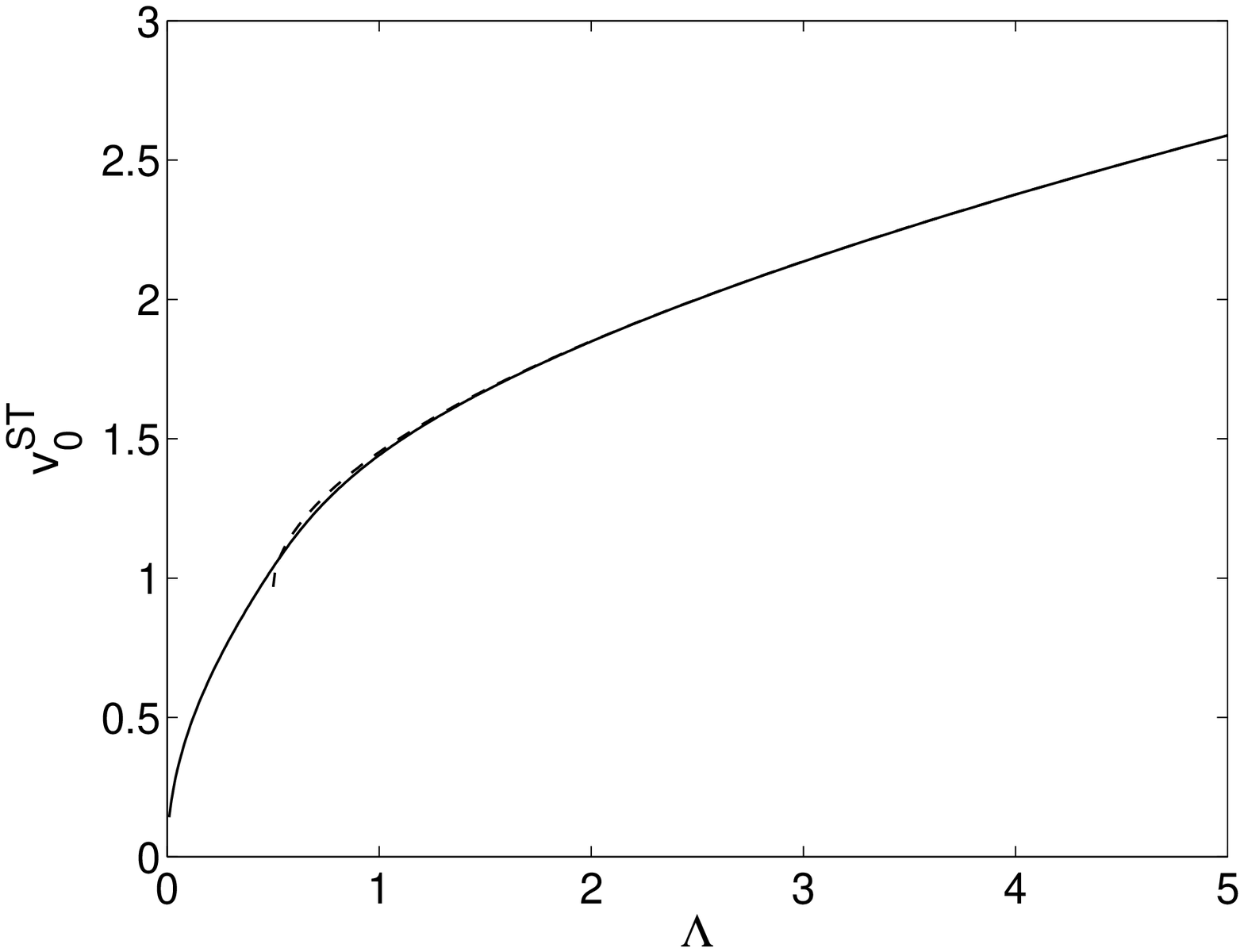} &
\includegraphics[width=\doublefig]{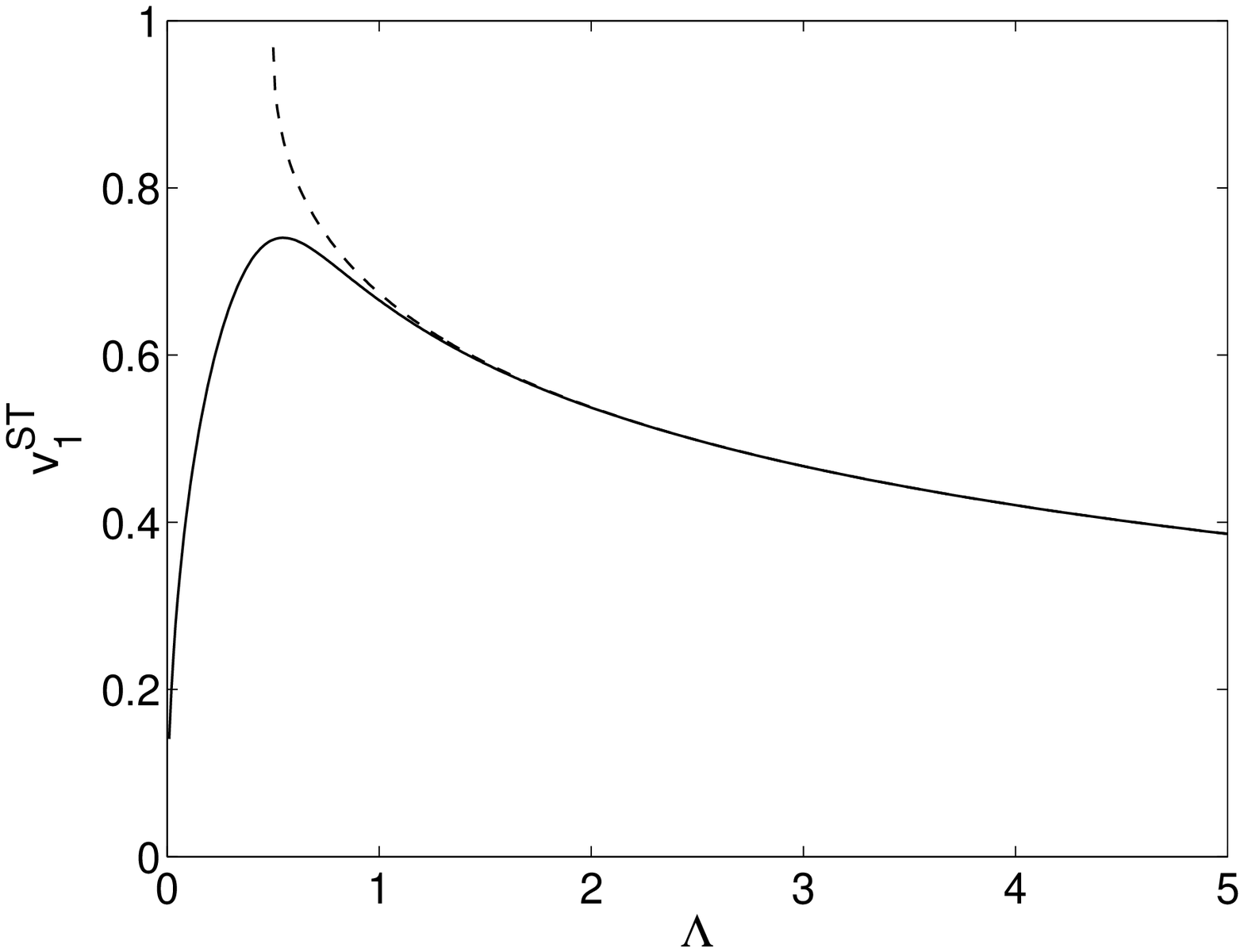}
\end{tabular}%
\end{center}
\caption{Same as Fig. \ref{fig:profvar1} but with dashed lines
corresponding to approximation (\ref{xST}).}
\label{fig:profhomo1}
\end{figure}
\begin{figure}
\begin{center}
\begin{tabular}{cc}
\includegraphics[width=\doublefig]{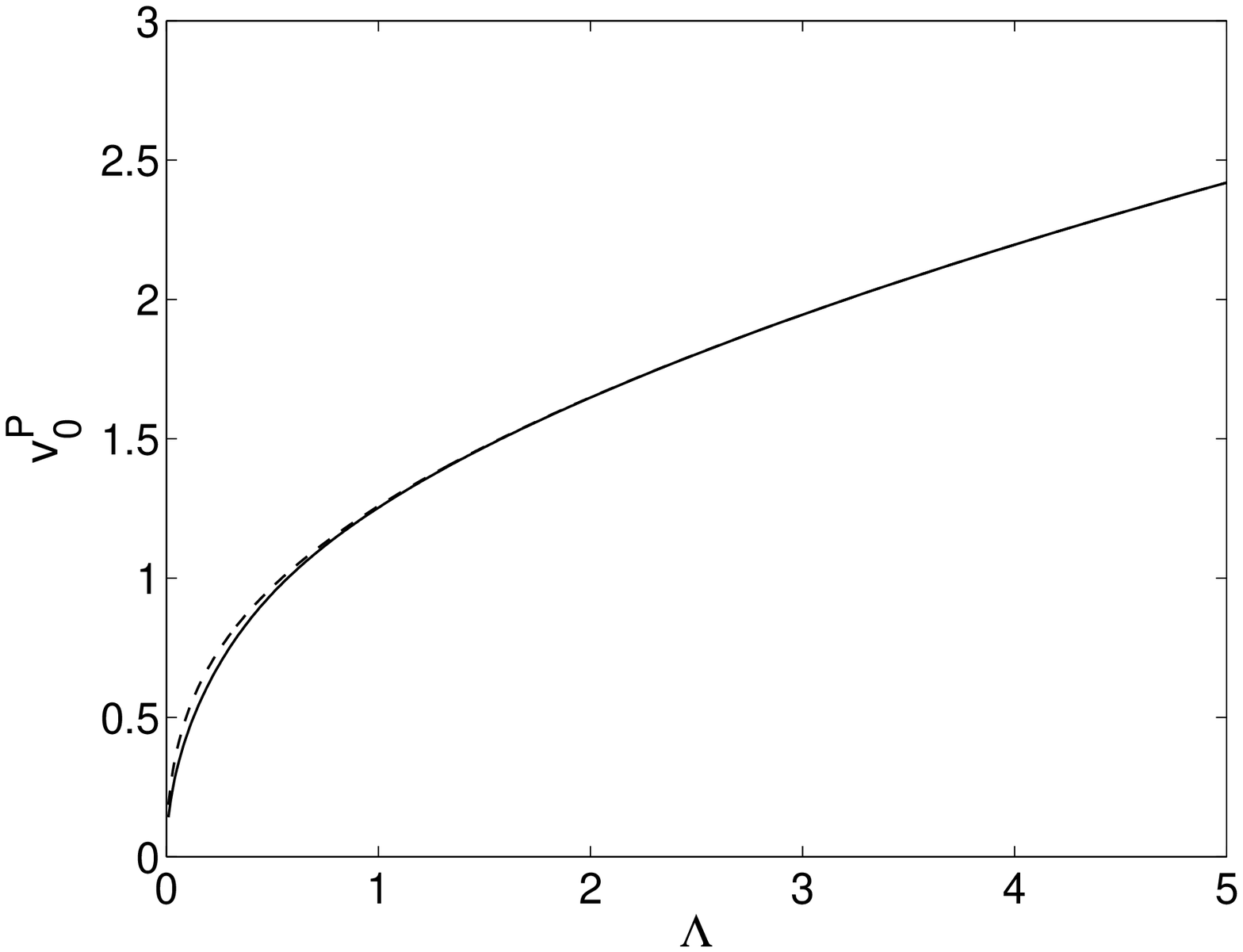} &
\includegraphics[width=\doublefig]{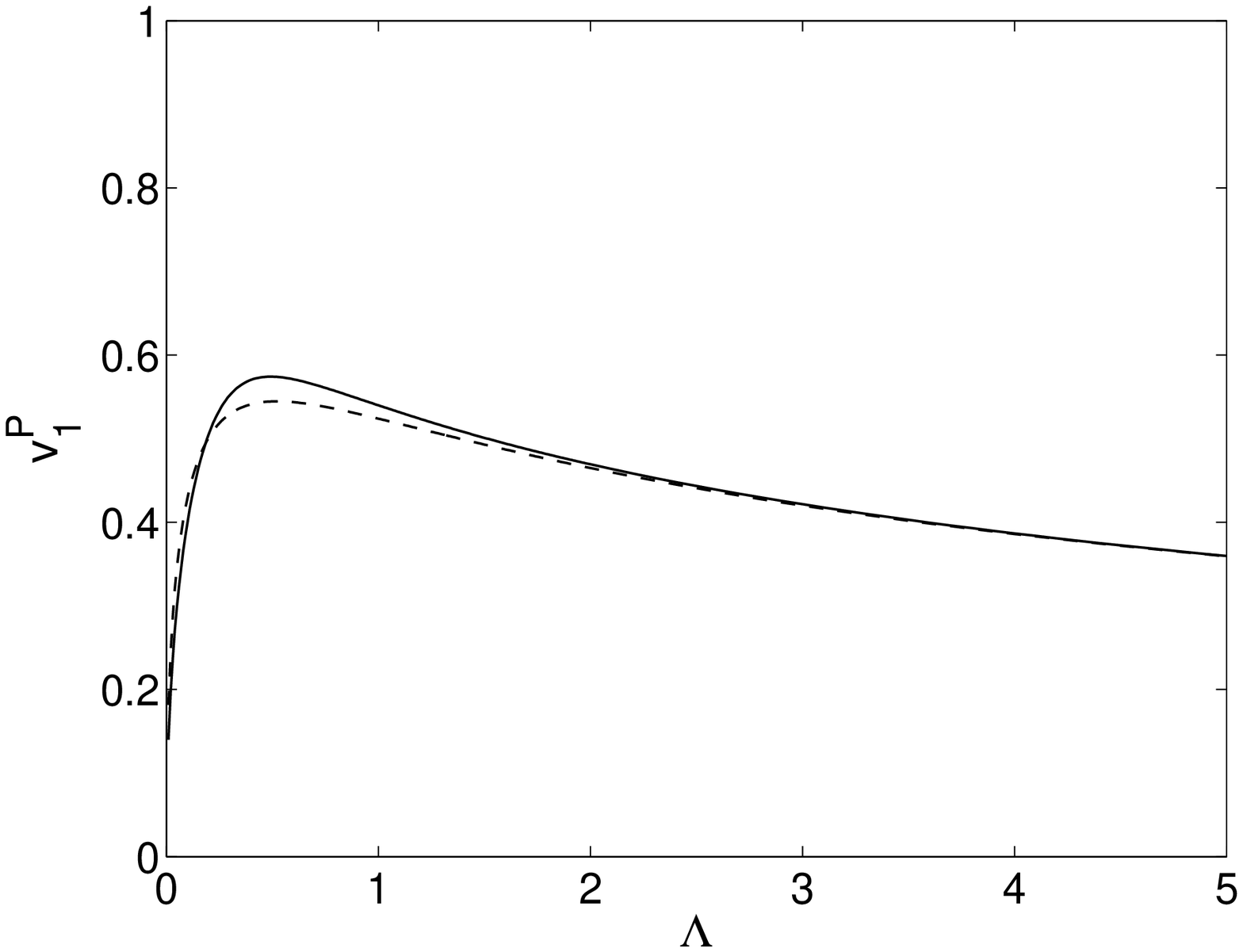}
\end{tabular}%
\end{center}
\caption{Same as Fig. \ref{fig:profvar2} but with dashed lines
corresponding to approximation (\ref{xPb}).}
\label{fig:profhomo2}
\end{figure}
This equation has nine solutions (see Fig.
\ref{fig:intersections}a). One of them ($x=0$), corresponds to the
origin. Once this solution is eliminated, the reminder equation is a
bi-quartic one. Thus, if $x=\xi$ is a solution of (\ref{bigeq}),
$x=-\xi$ is also a solution: this is due to the fact that $\pm v_n$
is a solution of (\ref{beq2}). Solutions $x=\xi_1$, $x=\xi_2$,
$x=\xi_0$ and $x=\xi_3$ in Fig. \ref{fig:intersections} correspond
to the positive solutions of (\ref{bigeq}). The point $x=\xi_0$ is
in the bisectrix of the first quadrant and corresponds to the P-mode
(i.e. $v_0^P=\xi_0$), and the point $x=\xi_3$ lies in the bisectrix
of the fourth quadrant and corresponds to a twisted mode (i.e. a
discrete soliton with two adjacent excited sites with the same
amplitude and opposite sign). Setting $y(\xi_0)=\xi_0$ and
$y(\xi_3)=-\xi_3$ in (\ref{unstable}), one obtains
$\xi_0=\lambda^{-2}\sqrt{(\lambda-1)(\lambda^4-1)}$,
$\xi_3=\lambda^{-2}\sqrt{(\lambda+1)(\lambda^4-1)}$.

Upon elimination of the roots $x=\xi_0$ and $x=\xi_3$ from
(\ref{bigeq}), $\xi_1$ and $\xi_2$ can be calculated as solutions of
a quadratic equation. Thus,
\begin{equation}\label{xST}
    \xi_1=\lambda^{-2}\sqrt{(\lambda^4-1)(\lambda-\sqrt{\lambda^2-4})/2}, \quad
    \xi_2=\lambda^{-2}\sqrt{(\lambda^4-1)(\lambda+\sqrt{\lambda^2-4})/2}.
\end{equation}
These solutions are related with the ST-mode as $v_0^{ST}=\xi_2$ and
$v_1^{ST}=\xi_1$. On the other hand, for the P-mode,
$v_0^{P}=\xi_0$, and, $v_1^{P}$ should be determined by application
of the map (\ref{map}). This yields
\begin{equation}\label{xPb}
v_0^{P}=\lambda^{-2}\sqrt{(\lambda-1)(\lambda^4-1)},\quad
v_1^P=\lambda^{-6}(\lambda^3+\lambda-1)\sqrt{(\lambda-1)(\lambda^4-1)}.
\end{equation}
In Figs. \ref{fig:profhomo1} and \ref{fig:profhomo2}, the values of
$v_0$ and $v_1$ obtained through the homoclinic approximation are
represented versus $\Lambda$ and compared with the exact numerical
results. It can be observed that, for ST-modes, no approximate
solutions exist for $\Lambda<0.5$. For $\Lambda=1/2$ (i.e.
$\lambda=2$), the points $(\xi_1,\xi_2)$ and $(\xi_2,\xi_1)$
disappear via a pitchfork bifurcation at $(\xi_0,\xi_0)$ (see Fig.
\ref{fig:intersections}b). This artifact is a by-product of
the decreasing accuracy of our approximations as $\Lambda \rightarrow 0$;
as discussed before, the ST-mode should exist for all values of $\Lambda > 0$.

\subsection{The Sievers--Takeno approximation}

\begin{figure}
\begin{center}
\begin{tabular}{cc}
\includegraphics[width=\doublefig]{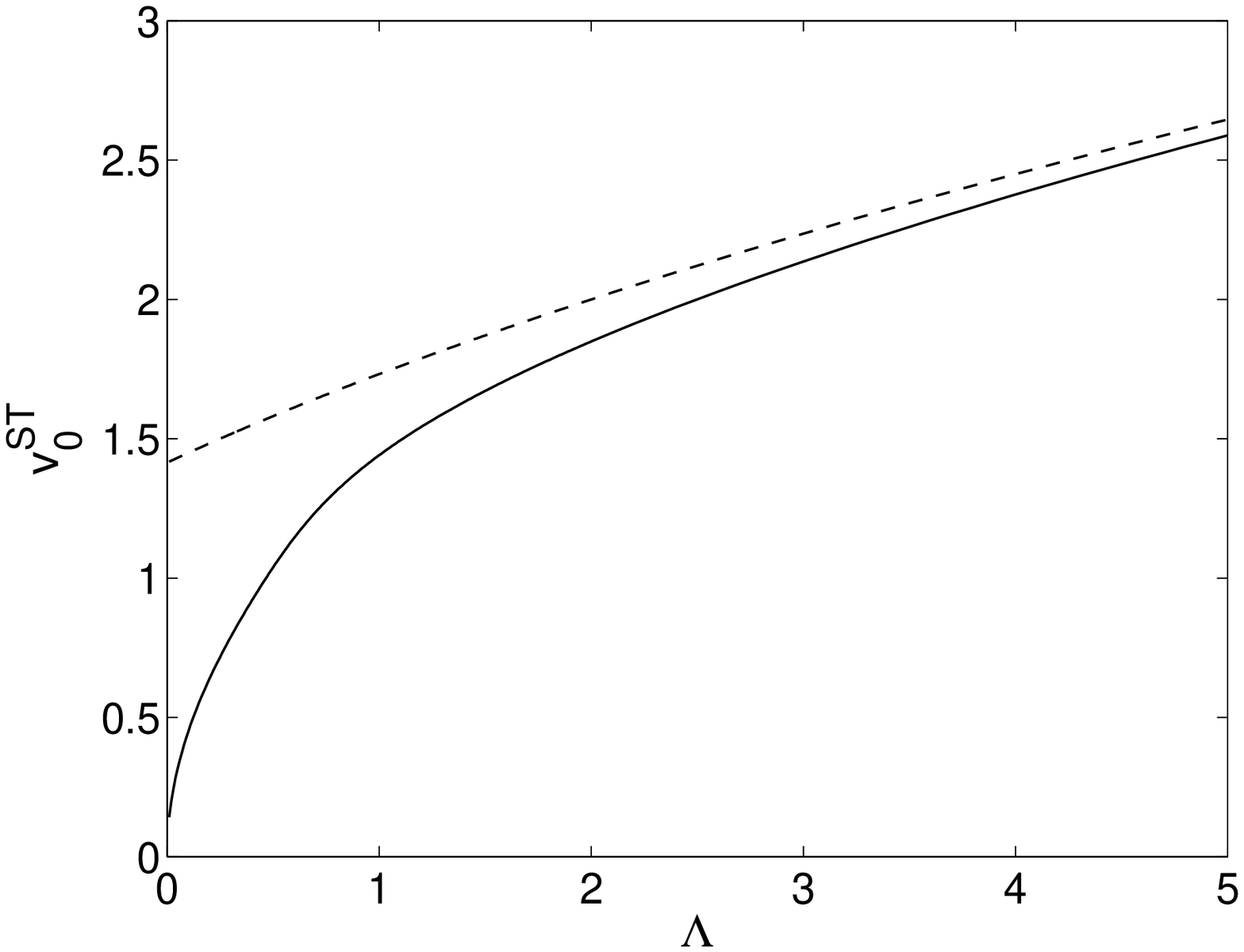} &
\includegraphics[width=\doublefig]{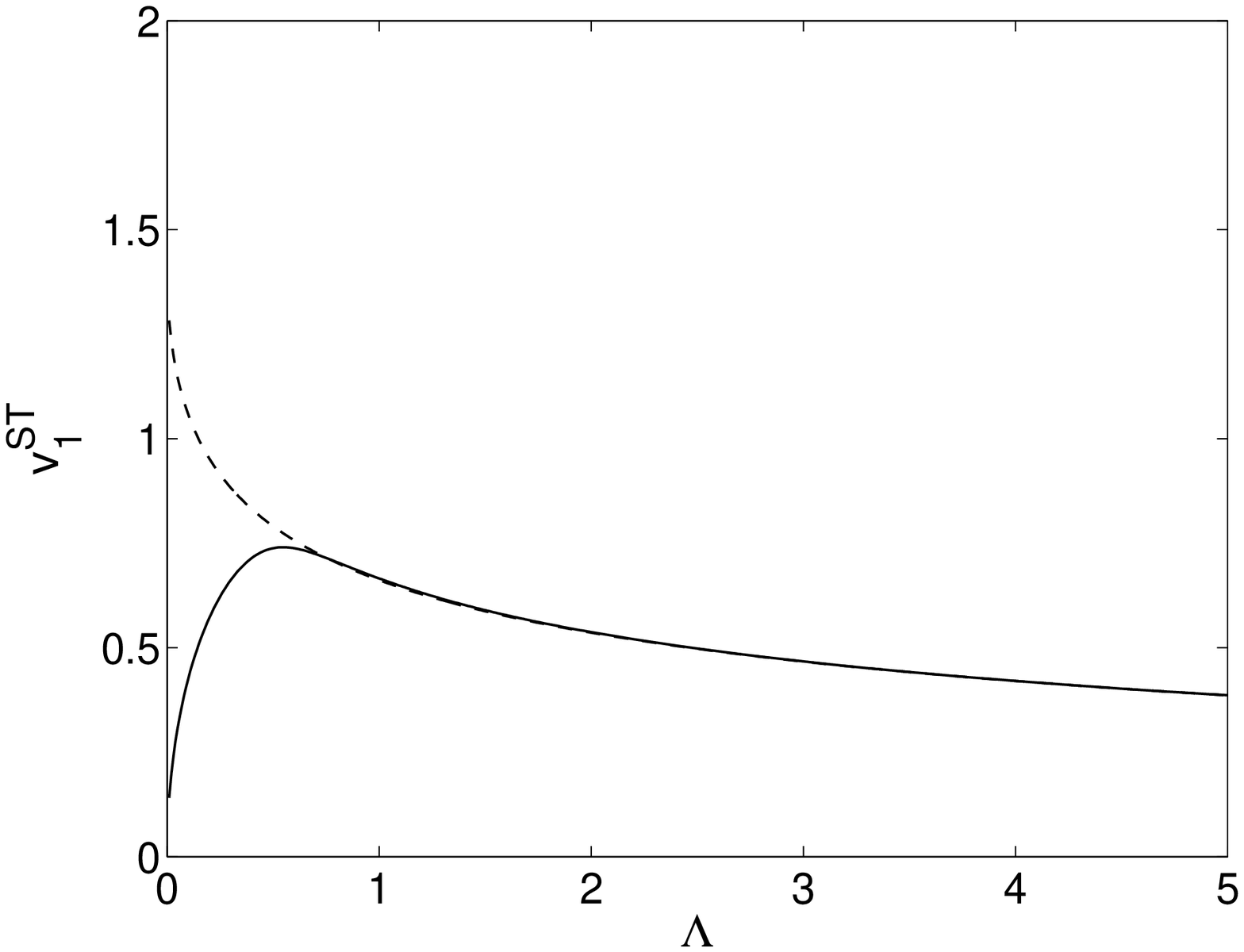}
\end{tabular}%
\end{center}
\caption{Same as Fig. \ref{fig:profvar1} but with dashed lines
corresponding to approximation (\ref{appv}).}
\label{fig:Takeno}
\end{figure}
A method to approximate solutions of (\ref{beq2}) has been
introduced by Sievers and Takeno, for a recurrence relation similar
to it but with slightly different nonlinear terms \cite{ST}. This
approach has been generalized to the $d$-dimensional DNLS equation
in reference \cite{Tak89}. In what follows we briefly describe the
method, incorporating some precisions and simplifications. Setting
$v_n = v_0 \eta_n$, equation (\ref{beq2}) becomes
\begin{equation}
\label{rec2} \eta_{n+1}-2 \eta_n +\eta_{n-1}=\Lambda \eta_n -v_0^2
\eta_n^3 ,
\end{equation}
with $\eta_{-n}=\eta_n$, $\eta_0 =1$. Setting $n=0$ in (\ref{rec2})
we obtain in particular
\begin{equation}
\label{site1} v_0^2 = \Lambda + 2(1-\eta_1 ).
\end{equation}
Equation (\ref{rec2}) can be rewritten as a suitable nonlocal
equation using a lattice Green function in conjunction with the
reflectional symmetry of $\eta_n$ and equation (\ref{site1}). This
yields for all $n\geq 1$
\begin{equation}
\label{nonloc} \eta_n = [ \, \Lambda + 2(1-\eta_1)\, ]
\frac{\lambda^{-n}}{\lambda - \lambda^{-1}} +\sum_{k\geq
1}{\eta_k^3\, (\lambda^{-|n-k|}+\lambda^{-n-k})},
\end{equation}
where $\lambda\equiv\lambda_+$ is given by (\ref{eigen}). Problem
(\ref{nonloc}) can be seen as a fixed point equation $\{ \eta \} =
F_{\Lambda}(\{ \eta \})$ in $\ell_\infty (\mathbb{N}^\ast)$. Noting
$B_\epsilon$ the ball $\| \{ \eta \} \|_{\ell_\infty
(\mathbb{N}^\ast)} \leq \epsilon$, the map $F_{\Lambda}$ is a
contraction on $B_\epsilon$ provided $\epsilon$ is sufficiently
small and $\Lambda$ is greater than some constant $\Lambda_0
(\epsilon )$. In that case, the solution of (\ref{nonloc}) is unique
in $B_\epsilon$ by virtue of the contraction mapping theorem and it
can be computed iteratively. Choosing $\{ \eta \}=0$ as an initial
condition, we obtain the approximate solution
\begin{equation}
\label{appet} \eta_n \approx (\, F_{\Lambda}(0)\, )_n =
\frac{\Lambda + 2}{\lambda - \lambda^{-1}}\, \lambda^{-n} , \ \ \
n\geq 1.
\end{equation}
Obviously the quality of the approximation would increase with
further iterations of $F_{\Lambda}$.
Using (\ref{appet}) and (\ref{site1})
in the limit when $\Lambda$ is large, we obtain
\begin{equation}
\label{appv} v_n \approx   (\Lambda + 2)^{1/2}\, \lambda^{-|n|}
\end{equation}
since $\lambda \sim \Lambda $ as $\lambda \rightarrow +\infty$. The
values of $v_0$ and $v_1$ in this approximation are compared with
the exact numerical results in Fig. \ref{fig:Takeno}. We observe
that the approximation captures the asymptotic behaviour of $v_0$
and $v_1$ for $\Lambda\rightarrow\infty$.

\section{The quasi-continuum approximation}

\begin{figure}
\begin{center}
\begin{tabular}{cc}
\includegraphics[width=\smalldoublefig]{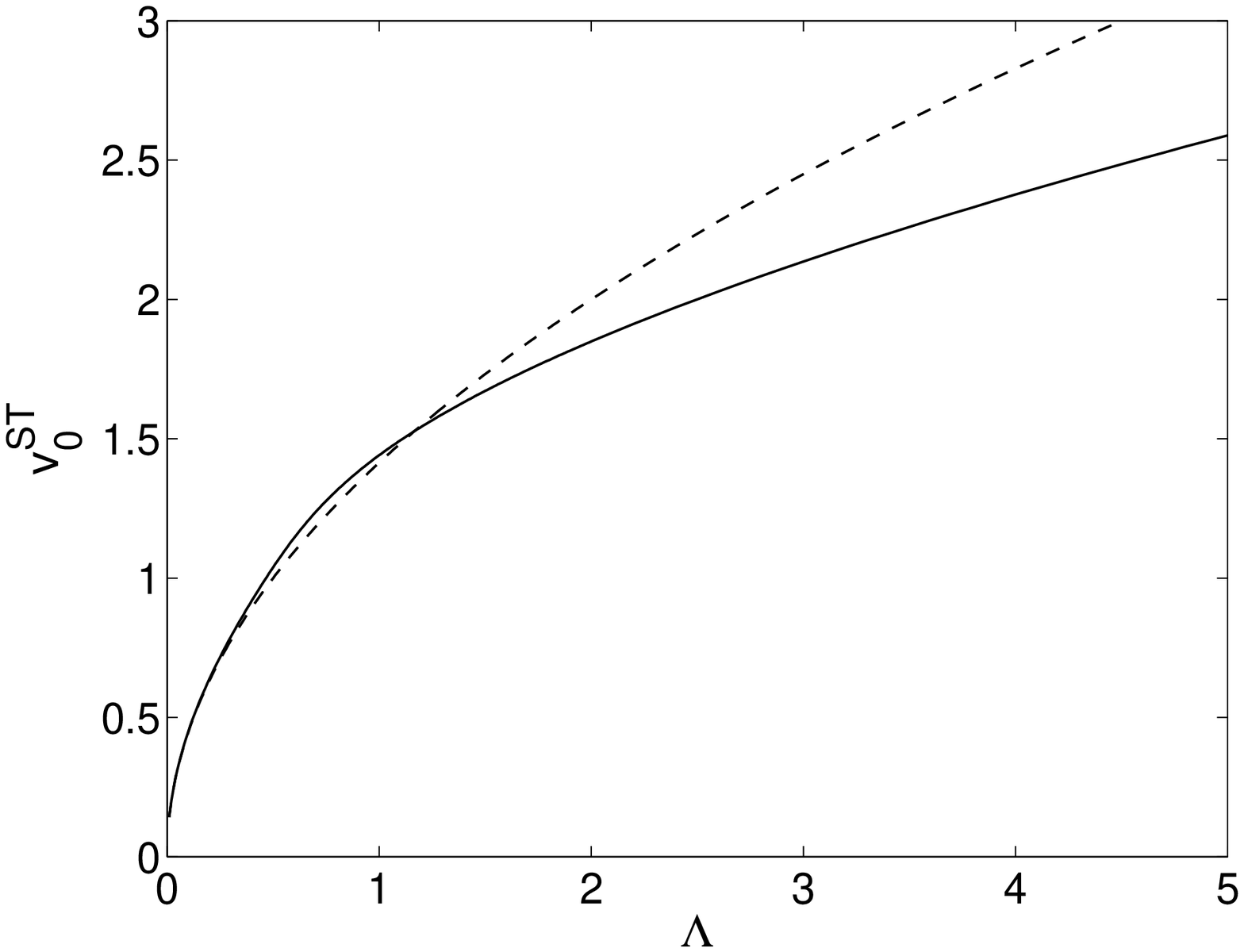} &
\includegraphics[width=\smalldoublefig]{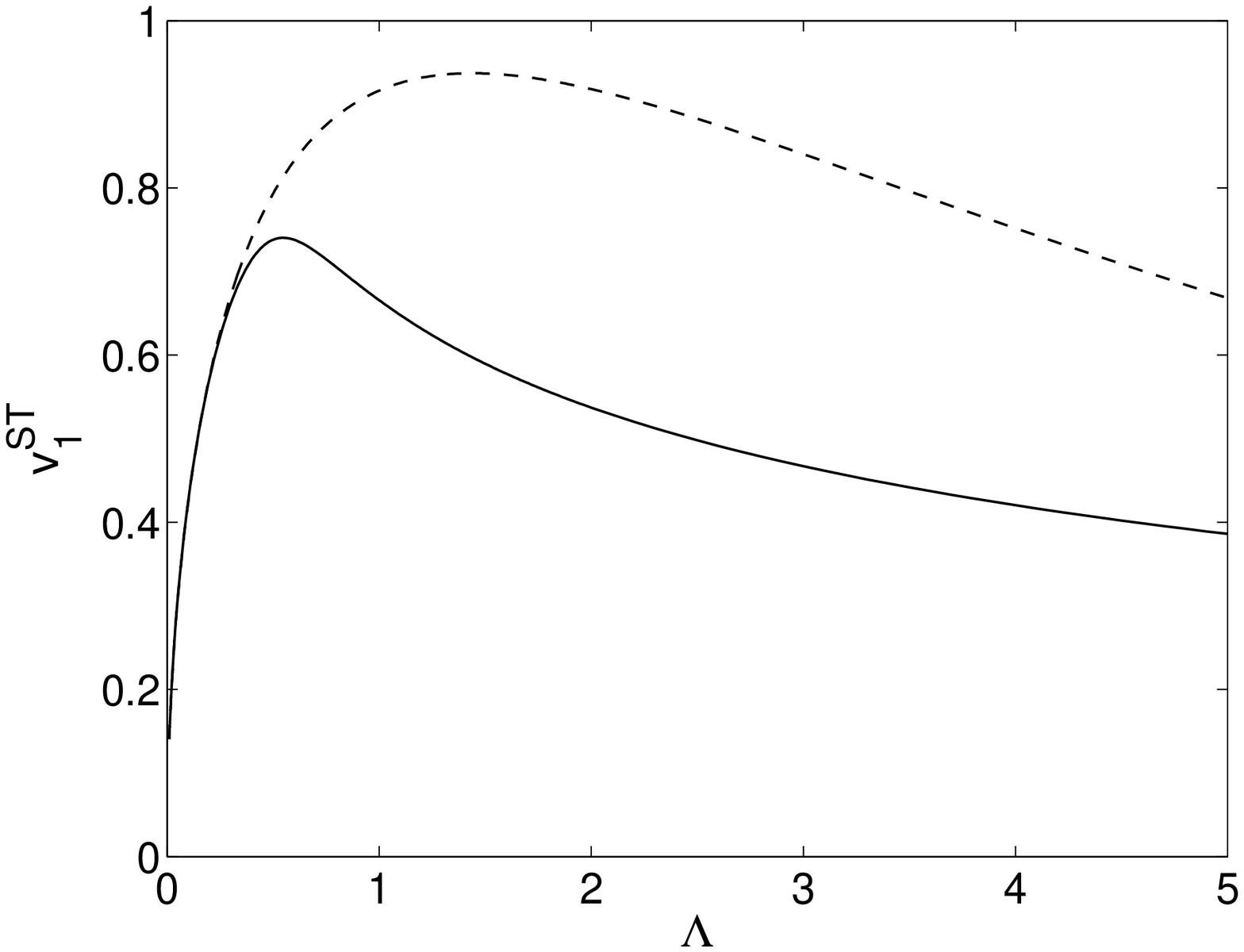}
\end{tabular}%
\end{center}
\caption{Same as Fig. \ref{fig:profvar1} but with dashed lines
corresponding to approximation (\ref{ceq1}).}
\label{fig:profcont1}
\end{figure}
\begin{figure}
\begin{center}
\begin{tabular}{cc}
\includegraphics[width=\smalldoublefig]{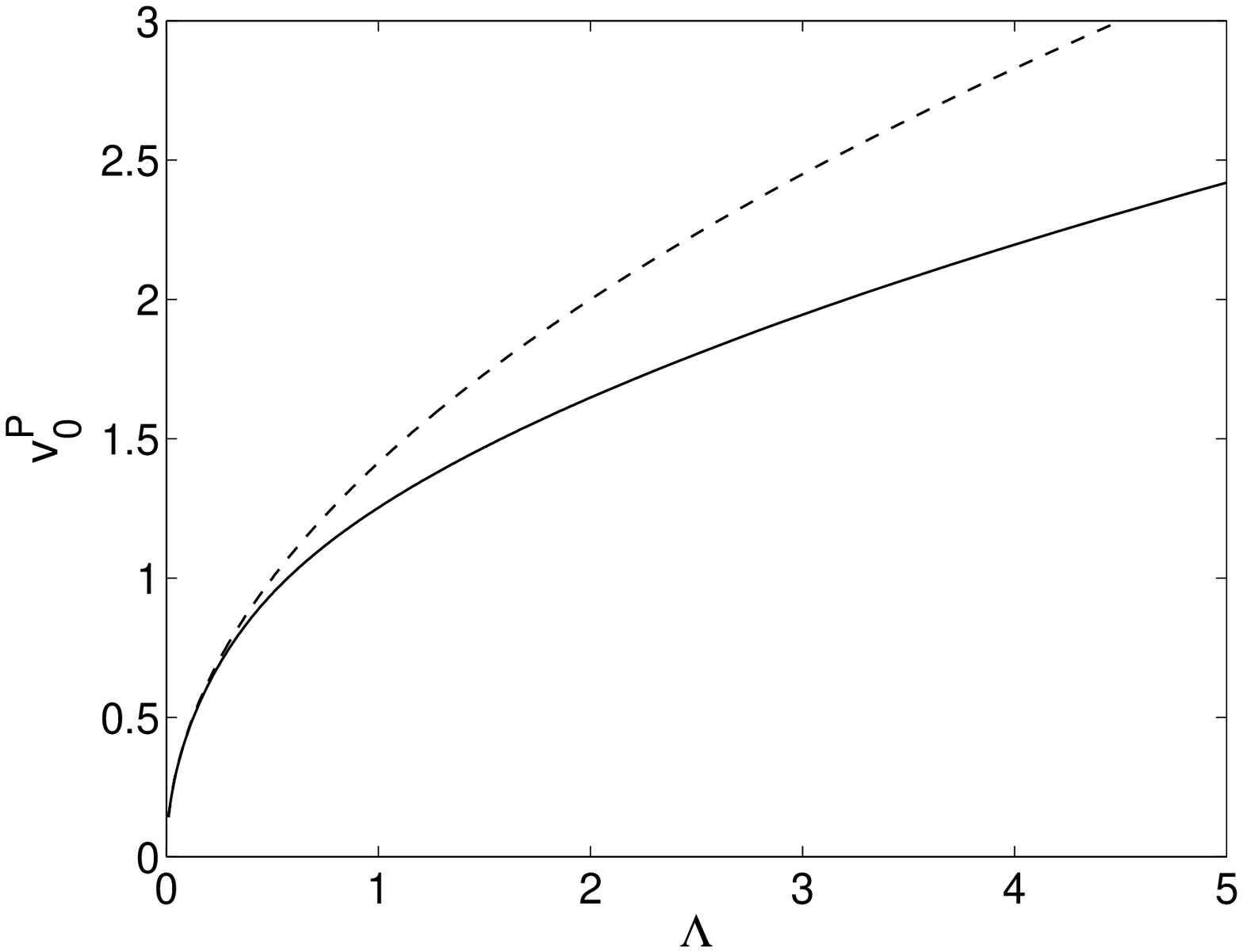} &
\includegraphics[width=\smalldoublefig]{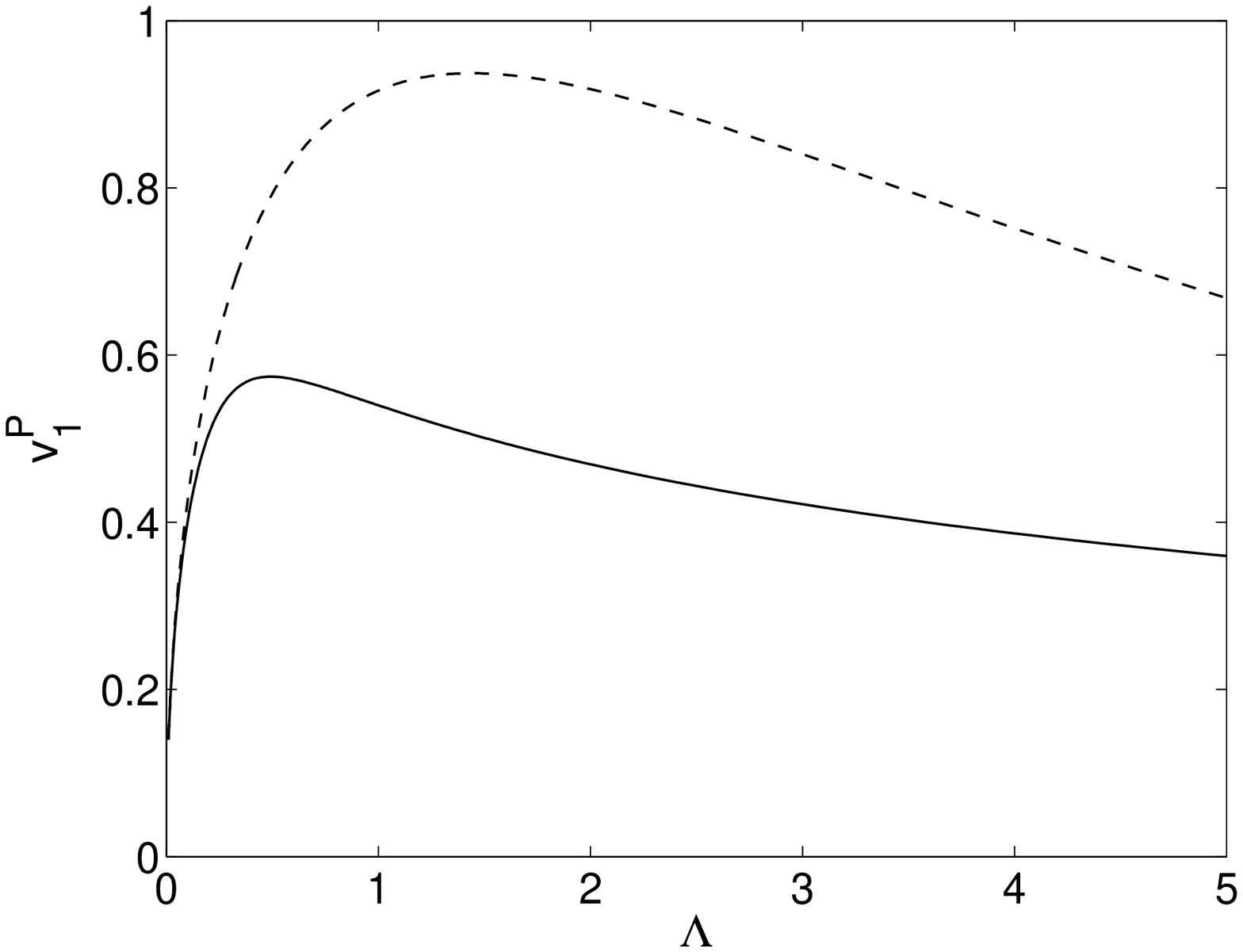}
\end{tabular}%
\end{center}
\caption{Same as Fig. \ref{fig:profvar2} but with dashed lines
corresponding to approximation (\ref{ceq1}).}
\label{fig:profcont2}
\end{figure}
As it can be concluded from previous sections, none of the
established approximations perform well for $\Lambda$ close to zero
(although the VA is notably more accurate than the invariant
manifold and Sievers--Takeno approximation). A quasi-continuum
approximation could be used to fill this gap. To this end, we follow
Eqs. (13) and (14) of Ref. \cite{PRB}. Then the ST- and P-modes can
be approximated by the continuum soliton based expressions:
\begin{equation}
v^{ST}_{n}=\sqrt{2\Lambda}\sech(n\sqrt{\Lambda}), \qquad
v^{P}_{n}=\sqrt{2\Lambda}\sech[(|n+1/2|-1/2)\sqrt{\Lambda}].
\label{ceq1}
\end{equation}
These expressions lead to the results shown in Figs.
\ref{fig:profcont1} and \ref{fig:profcont2}. Naturally, this
approach captures the asymptotic limit $v_0\sim\sqrt{2\Lambda}$ when
$\Lambda\rightarrow0$, but fails increasingly as $\Lambda$ grows.

\section{Summary and conclusions}

\begin{figure}
\begin{center}
\begin{tabular}{cc}
\includegraphics[width=\doublefig]{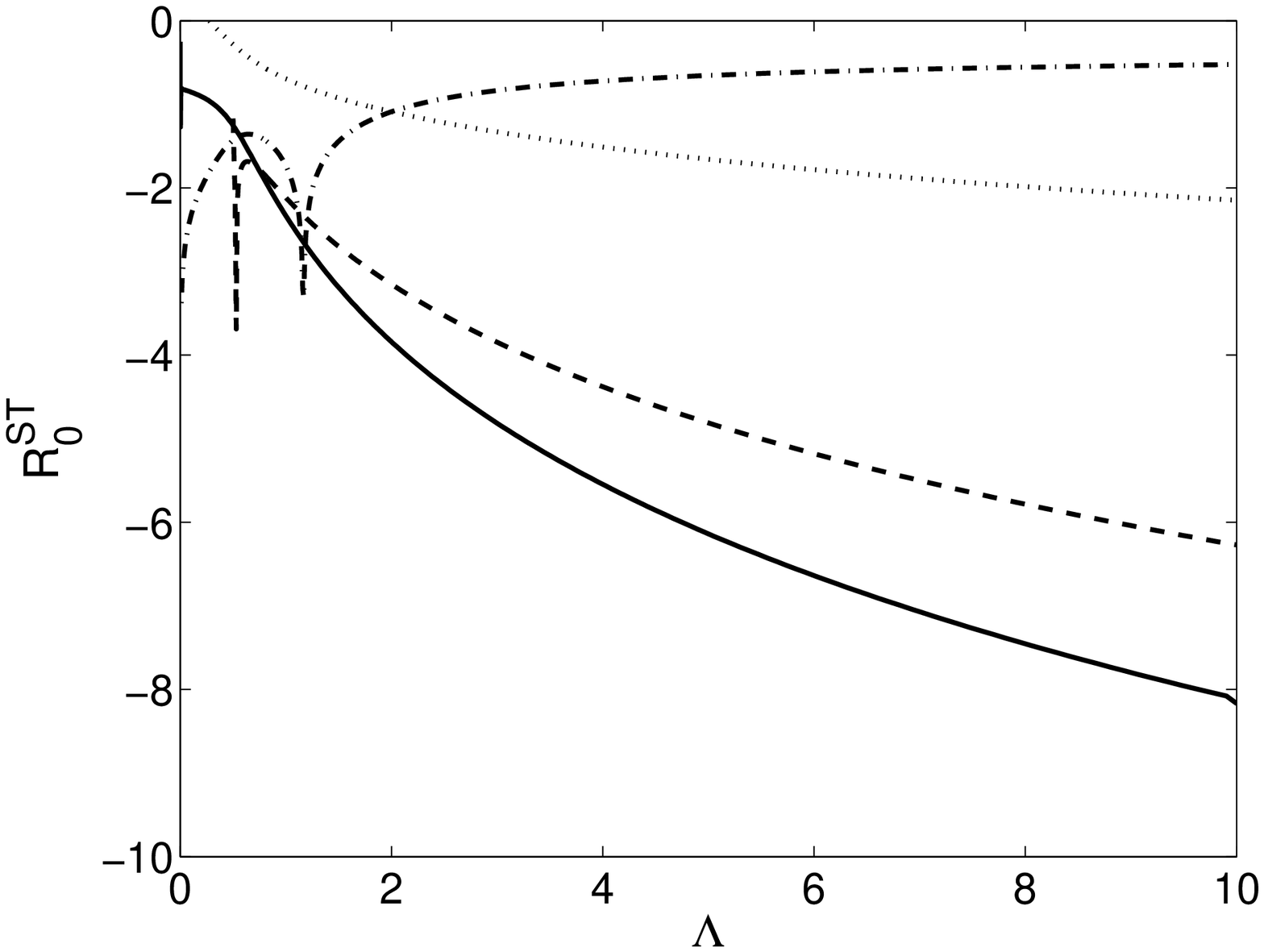} &
\includegraphics[width=\doublefig]{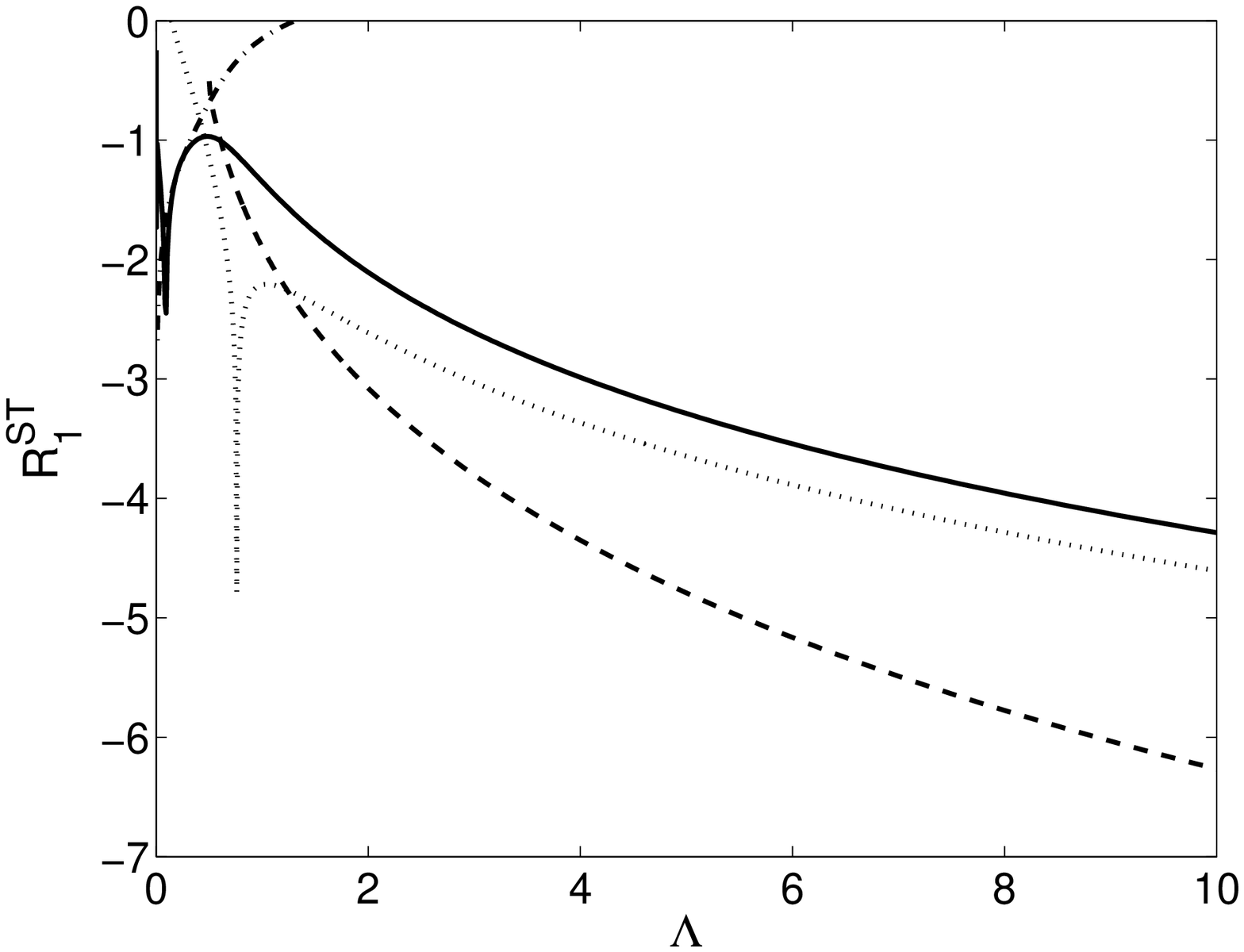}
\end{tabular}%
\end{center}
\caption{Representation of variable $R$ defined in (\ref{R}) versus
$\Lambda$ for ST-modes. Full lines correspond to the variational
approach; the dashed line corresponds to the homoclinic
approximation; the dash-dotted lines to the continuum approximation;
and the dotted line to the Sievers--Takeno approximation.}
\label{fig:compare1}
\end{figure}

\begin{figure}
\begin{center}
\begin{tabular}{cc}
\includegraphics[width=\doublefig]{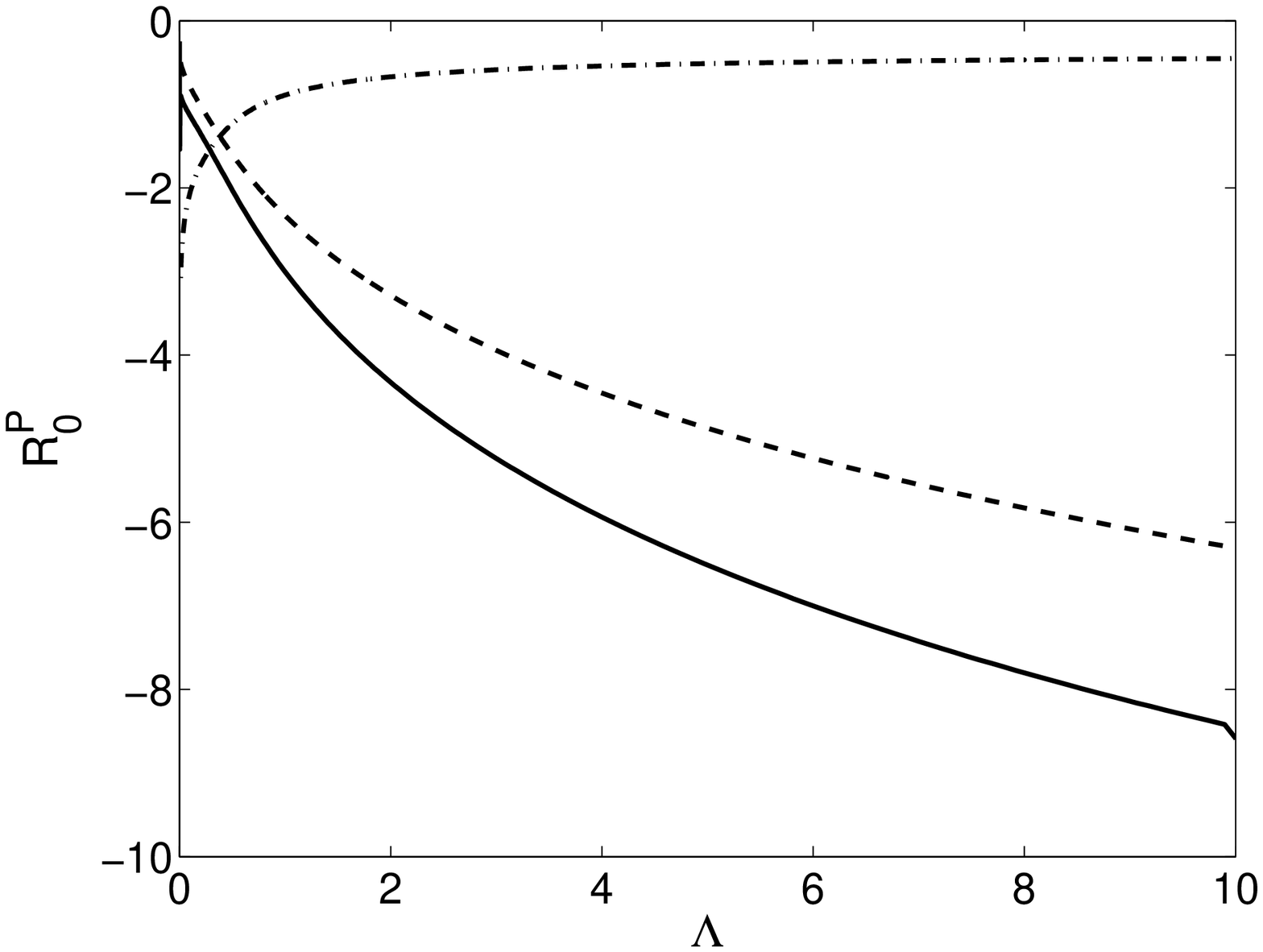} &
\includegraphics[width=\doublefig]{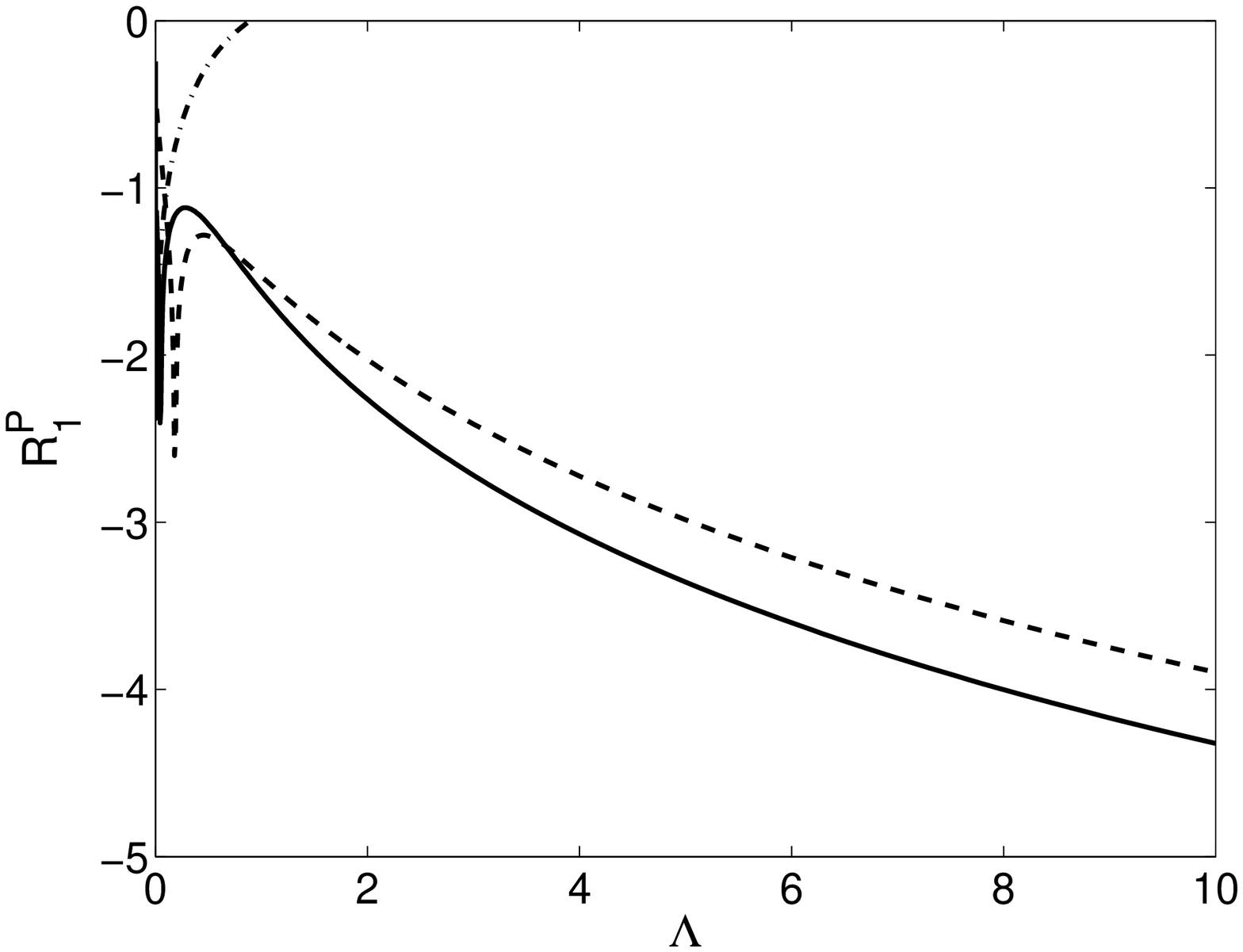}
\end{tabular}%
\end{center}
\caption{Representation of variable $R$ defined in (\ref{R}) versus
$\Lambda$ for P-modes. Full lines corresponds to variational
approach; dashed line, to the homoclinic approximation; and
dash-dotted lines, to the continuum approximation.}
\label{fig:compare2}
\end{figure}

In Figs. \ref{fig:compare1} and \ref{fig:compare2} the results of
the paper are summarized. To this end, a variable, giving the
relative error at site $n$, is defined as:
\begin{equation}\label{R}
    R_n=\log_{10}\left|(v_n^{\mathrm{approx}}-v_n^{\mathrm{exact}})
    /v_n^{\mathrm{exact}}\right|.
\end{equation}
We can generally conclude that the variational approximation offers
the most accurate representation of the amplitude amplitude of the
Page mode $v_n^P$ at the two sites $n=0$ and $n=1$ with some small
exceptions. These involve some particular intervals of $\Lambda$
where the homoclinic approximation may be better and also the
interval sufficiently close to the continuum limit, where the best
approximation is given by the discretization of the continuum
solution.  Similar features are observed for the approximation of
the Sievers--Takeno mode $v_n^{ST}$ at site $n=0$. However, a
different scenario occurs for this mode at site $n=1$, since the
homoclinic approximation gives the best result for $\Lambda > 1.5$.
As $\Lambda$ goes to $0$, the Sievers-Takeno, variational and
quasi-continuum approximations give successively the best results in
small windows of the parameter $\Lambda$. Notice that in the
interval $\Lambda\in(0,0.5]$ neither the variational, nor the
homoclinic approximation are entirely satisfactory. The latter
suffers, among other things, the serious problem of producing a
spurious bifurcation of two ST modes with a P-mode. On the other
hand, for larger values of $\Lambda$ (i.e., for $\Lambda > 0.5$),
the quasi-continuum approach is the one that fails increasingly
becoming rather unsatisfactory, while the discrete approaches are
considerably more accurate, especially for $\Lambda > 2$, when their
relative error drops below $1 \%$ (with the exception of the
Sievers-Takeno approximation of $v_0^{ST}$, which only reaches this
precision for $\Lambda > 10$).

We hope that these results can be used as a guide for developing
sufficiently accurate analytical predictions in different parametric
regimes for such systems. It would naturally be of interest to
extend the present considerations to higher dimensions. However, it
should be acknowledged that in the latter setting the variational
approach would extend rather straightforwardly, while the homoclinic
approximation is restricted to one space dimension and the other
approximations would become more technical.

\begin{acknowledgments}
JC and BSR acknowledge financial support from the MECD project
FIS2004-01183. PGK gratefully acknowledges support from NSF-CAREER,
NSF-DMS-0505663 and NSF-DMS-0619492. We acknowledge F Palmero for
his useful comments.
\end{acknowledgments}

\label{lastpage}

\end{document}